\documentclass[showpacs,twocolumn,aps,prx,floatfix,superscriptaddress]{revtex4-1}
\usepackage{amsmath,amssymb,graphicx,bm,color,mathrsfs,verbatim,caption}

\pdfoutput=1

\newcommand{\be}{\begin{equation}}
\newcommand{\ee}{\end{equation}}
\newcommand{\bea}{\begin{eqnarray}}
\newcommand{\eea}{\end{eqnarray}}
\begin{document}

%\title{Dynamical preparation of Floquet Chern insulators: \\
%A no-go theorem and the experiments}

\title{Dynamical preparation of Floquet Chern insulators}

\author{Luca D'Alessio}
\affiliation{Department of Physics, The Pennsylvania State University, 
University Park, PA 16802, USA}
\affiliation{Department of Physics, Boston University, Boston, MA 02215, USA}

\author{Marcos Rigol}
\affiliation{Department of Physics, The Pennsylvania State University, 
University Park, PA 16802, USA}

\begin{abstract}
Realizing topological insulators is of great current interest because of their remarkable properties and possible future applications. There are recent proposals, based on Floquet analyses, that one can generate topologically nontrivial insulators by periodically driving topologically trivial ones. Here we address what happens if one follows the dynamics in such systems. Specifically, we present an exact study of the time evolution of a graphene-like system subjected to a circularly polarized electric field. We prove that, for infinite (translationally invariant) systems, the Chern number is conserved under unitary evolution. For systems with boundaries, on the other hand, we show that a properly defined topological invariant, the Bott index, can change. Hence, it should be possible to experimentally prepare topological states starting from non-topological ones. We show that the chirality of the edge current in such systems can be controlled by adjusting the filling.
\end{abstract}
\maketitle

\section*{Introduction}

Topological and Chern Insulators are fascinating phases of quantum matter that are qualitatively different from conventional insulators and semiconductors \cite{RMP1,RMP2}. They are characterized by a gap in the bulk and have topologically protected gapless excitations at the boundary \cite{gapless1,gapless2}. Topological phases fall outside the Landau-Ginsburg (effective) theory of spontaneous symmetry breaking and are characterized by bulk topological invariants, such as the Chern number~\cite{chern}, which can be interpreted as non-local order parameters. Recently, it has been proposed that time-periodic perturbations can induce topological properties in otherwise non-topological materials, opening the exciting possibility of studying non-equilibrium topological transitions \cite{graphene0,graphene1,natanel}. 

The link between topology and time-periodic driving can be established via the Floquet theorem \cite{floquet,floquet1,floquet2,Reichl}, which is very similar to Bloch's theorem \cite{aschcroft}. The Floquet theorem states that the evolution operator of any system described by a time periodic Hamiltonian $H(t)=H(T+t)$ can be factorized as 
\be
{U}(t,0)={P}(t,0)\, e^{-i\, {H}_{F}[0]t/\hbar}
\label{eq:Floquet-theorem}
\ee
where ${P}(t,0)={P}(t+T,0)$ is a unitary periodic operator and ${H}_{F}[0]$ is the time-independent Floquet Hamiltonian. Being time-independent, the Floquet Hamiltonian can be characterized using standard concepts developed for undriven situations. For example, the Floquet Hamiltonian is considered topological if the Chern number of the Floquet bands is non-zero~\cite{natanel}. As noted in Ref.~\cite{anomalous_correspondence}, this characterization is incomplete since it ignores the properties encoded into the operator $P$. Moreover, periodically driven systems are manifestly out of equilibrium and the topological properties of the time-evolving state do not need to reflect the topology of the underlying Floquet Hamiltonian.

Here we extend the topological characterization above to isolated, thermodynamically large, out-of-equilibrium systems. In particular, we address what should happen in an experiment on an isolated system when one turns on the periodic driving using linear ramps. For infinite (translationally invariant) systems, in the absence of dissipation, we prove a no-go theorem. We show that the Chern number is conserved under unitary evolution. On the other hand, for systems with boundaries, we show that a properly defined topological invariant, the Bott index \cite{bott}, can change. Hence, it is possible to dynamically prepare a topological wavefunction starting from a non-topological one via unitary evolution.

\section*{Results}

\subsection*{Model}

We consider the following Hamiltonian (or a unitary equivalent):
\begin{equation}
{H}(t)={H}_{S}+f(t)\, {H}_{1}(t)\label{eq:def_H(t)}
\end{equation}
where ${H}_{S}$ is time-independent, ${H}_{1}(t)$ is time periodic with period $T$, and the amplitude $f(t)$ is given by 
\be
f(t)\equiv\left\{ \begin{array}{ccc}
0 & \text{for} & t\le0\\
t/\tau & \text{for} & 0<t<\tau\\
1 & \text{for} & t>\tau
\end{array}\right.
\ee
We restrict our analysis to noninteracting fermionic Hamiltonians, for which a complete characterization of the (equilibrium) topological phases exists~\cite{kitaev}. We take the initial state $|\psi_0^{S}\rangle$ to be the ground state of the static Hamiltonian ${H}_{S}$. At time $t>0$, the time-dependent wavefunction is $|\psi(t)\rangle$. We are interested in situations in which the undriven system is described by a topologically trivial Hamiltonian ${H}_{S}$ and the driving is such that the Floquet Hamiltonian ${H}_{F}$ is topologically non-trivial.

We focus on spinless fermions in a honeycomb lattice with nearest-neighbor hopping $J$ and a staggered sublattice potential $\Delta$ subjected to a circularly polarized electric field $\vec{E}_\text{ac}(t)=E_{0}\, f(t)\,\left[-\cos\left(\Omega\,t\right)\,,\,\sin\left(\Omega\,t\right)\right]$ \cite{graphene0,graphene1}. In the electromagnetic gauge, in which the vector potential is zero, the time-dependent Hamiltonian is given by Eq.~\eqref{eq:def_H(t)} with
\bea
&&{H}_{S}=-J\sum_{\langle i\alpha,j\alpha'\rangle}
\left({c}_{i,\alpha}^{\dagger}\, {c}^{}_{j,\alpha'}+\text{H.c.}\right)
+\frac{\Delta}{2}\sum_{i}\left({n}_{i,1}-{n}_{i,2}\right)\nonumber\\
&&{H}_{1}(t)=\sum_{i\alpha}\, U_{E}(\vec{r}_{i,\alpha},t)\, {n}_{i,\alpha} 
\label{eq:H}
\eea
where the sum in ${H}_{S}$ extends over nearest neighbor sites, $\alpha\in\left\{ 1,2\right\}$ indicates the sublattice $\mathcal{A}$ and $\mathcal{B}$, respectively, ${n}_{i,\alpha}={c}_{i,\alpha}^{\dagger}{c}^{}_{i,\alpha}$ are the site number operators, $U_{E}(\vec{r},t)=-e\,\vec{r}\cdot\vec{E}_\text{ac}(t)$ is the electrostatic potential energy, and $e$ is the electric charge. 

We work in the high-frequency limit in which the driving frequency is larger than the single-particle band-width \cite{Esslinger}, i.e., $\hbar\Omega>6\, J$. In this limit, there is no ambiguity in the ordering of the Floquet quasi-energies and therefore the ground state of the Floquet Hamiltonian is well defined. Moreover, in order to obtain a non-trivial high-frequency limit, we scale the electric field with the frequency of the driving~\cite{marin1}, $eaE_{0}\propto \hbar\Omega$ where $a$ is the lattice spacing. Our parameters are:
\begin{equation}
\frac{\Delta}{J}=0.15,\quad\frac{\hbar\Omega}{J}=7,\quad 
K\equiv\frac{e\, a\, E_{0}}{\hbar\Omega}=1\label{eq:parameters}
\end{equation}
and are chosen so that the (effective) Floquet Hamiltonian $H_{F}$ is topological. The staggered sublattice potential $\Delta$ is introduced to make direct connection with the experiment in Ref.~\cite{Esslinger}, and to ensure that the edge modes that are not topological in nature are gapped out. The period of the driving is $T=2\pi/\Omega$ and we consider ramping times $\tau/T\in [0-2000]$. We stress that this choice of parameters is relevant for the recent experimental realization of the Haldane model in cold atoms \cite{Esslinger}. In Ref.~\cite{Esslinger}, $\tau=20\,$ms and $1/T=\Omega/(2\pi)=4\,$kHz so that $\tau/T=80$. However, the loading procedure there was more complex than the linear ramp considered here. 

\subsection*{Translationally Invariant System}

\begin{figure}[!t]
\includegraphics[width=0.9\columnwidth]{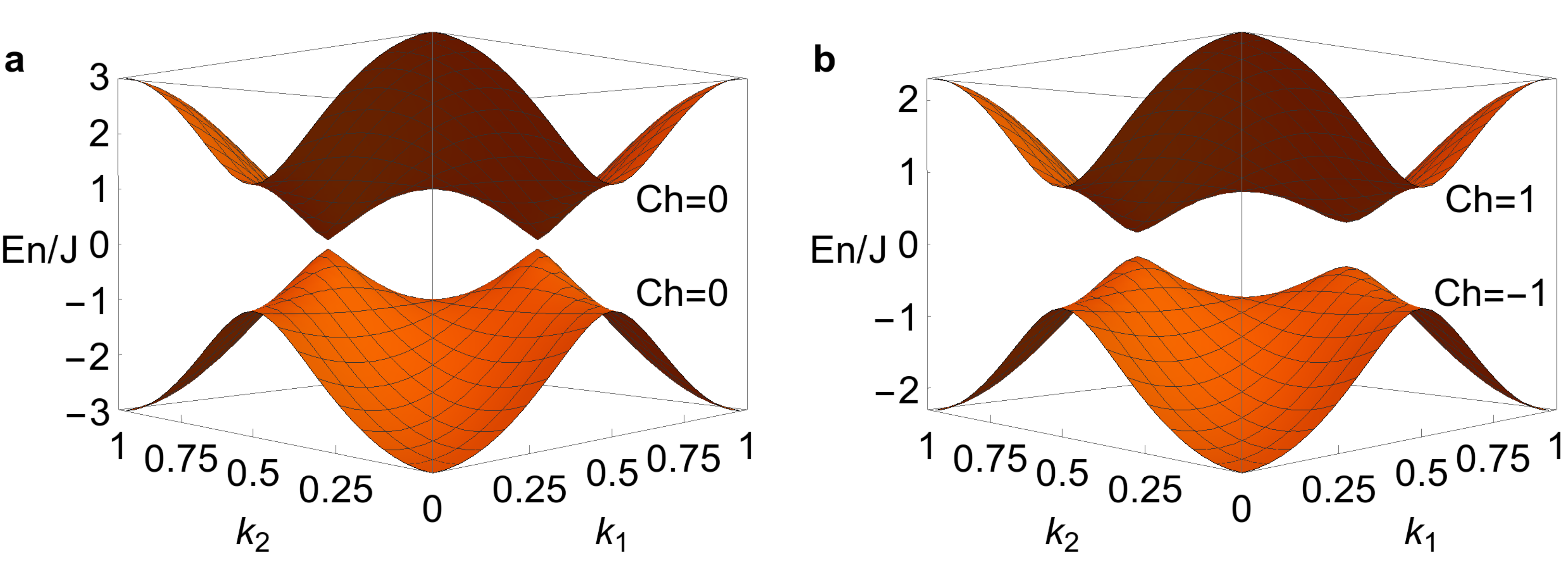}
\caption{Energy ``En" band structure of the infinite, translationally invariant, system in units of the hopping $J$. {\bf{a}}. The band structure of ${H}_{S}$ has a gap of size $\Delta$ at the two Dirac points and the Chern number of each band is zero.  {\bf{b}}. The band structure of ${H}_{F}$ has a gap of size $\sim0.30\, J$ and $\sim0.16\, J$ at the two Dirac points (the two gaps become equal only when $\Delta=0$), and the Chern number of the bands is $+1$ (top) and $-1$ (bottom). Moreover the bandwidth is renormalized from $6J$ to $6J\,\mathcal{J}_{0}(K)\approx4.59\, J$ (where $\mathcal{J}_{0}$ is the zeroth Bessel function of first kind and $K$ is defined in Eq.~\eqref{eq:parameters}).\label{fig:infinite}}
\end{figure}

We first consider the translationally invariant (infinite) system. In this case, it is convenient to work in the electromagnetic gauge in which the electric field is represented via the vector potential, i.e., ${\bf E}(t) = - \partial_t {\bf A}(t)$, as this gauge choice does not break translational invariance. By going to momentum space the system can be mapped, at half-filling, onto a collection of independent pseudo spin-$\frac{1}{2}$. The Hamiltonian ${H}=\sum_{\mathbf{k}}\, {H}_{\mathbf{k}}$ and the density matrix ${\rho}=|\psi\rangle\langle\psi|=\prod_{\mathbf{k}}{\rho}_{\mathbf{k}}$ are (we take $\hbar=1$ in what follows): 
\be
H_{\mathbf{k}}=-\frac{1}{2}\,\left(\mathbf{B}_{\mathbf{k}}
\cdot\mathbf{\mathbf{\sigma}}_{\mathbf{k}}\right),
\quad\rho_{\mathbf{k}}=\left(\frac{1_{2\times2}+
\mathbf{S}_{\mathbf{k}}\cdot\mathbf{\mathbf{\sigma}_{k}}}{2}\right)\label{eq:mapping}
\ee
Here, $1_{2\times2}$ is the $2\times2$ identity matrix, $\mathbf{\mathbf{\sigma}}$ are the Pauli matrices, and $\mathbf{S}_{\mathbf{k}}$ and $\mathbf{B}_{\mathbf{k}}$ are three dimensional, time-dependent, vectors fields defined in the two-dimensional Brillouin Zone (BZ) (see Fig.~\ref{fig:infinite}{\bf{a}}). For a pure state, the vector $\mathbf{S}_{\mathbf{k}}$ has unit length and the Chern number ($\text{Ch}$) of the state is simply the number of wrappings of the pseudo spin configuration around the Bloch sphere~\cite{chern}:
\begin{equation}
\text{Ch}(t)=\frac{1}{4\pi}\iint_{BZ}\, dk_{x}dk_{y}\,\mathbf{S}(t)\cdot
\left[\partial_{k_{x}}\mathbf{S}_{\mathbf{k}}(t)\times\partial_{k_{y}}
\mathbf{S}_{\mathbf{k}}(t)\right]\label{eq:chern}
\end{equation}
Here the integral extends over the BZ. In the ground state, the pseudo spin configuration is parallel to the pseudo magnetic field, i.e., $\mathbf{S}_{\mathbf{k}}=\mathbf{B}_{\mathbf{k}}/|\mathbf{B}_{\mathbf{k}}|$. This does not need to be the case out of equilibrium, where $\mathbf{S}_{\mathbf{k}}$ and $\mathbf{B}_{\mathbf{k}}$ are in general not parallel to each other. The exact equation of motion is:
\begin{equation}
i\partial_{t}\rho_{\mathbf{k}}=\left[H_{\mathbf{k}},
\rho_{\mathbf{k}}\right]\rightarrow\partial_{t}
\mathbf{S}_{\mathbf{k}}=\mathbf{S}_{\mathbf{k}}\times
\mathbf{B}_{\mathbf{k}},\label{eq:equation-motion}
\end{equation}
which is simply the precession of the pseudo spin $\mathbf{S}_{\mathbf{k}}$ around the pseudo magnetic field $\mathbf{B}_{\mathbf{k}}$.

With this mapping, the ground states $|\psi_0^{S}\rangle$ and $|\psi_0^{F}\rangle$ obtained by filling the valence bands of ${H}_{S}$ and ${H}_{F}$ are represented by the pseudo spin configurations $\mathbf{S}_{\mathbf{k}}^{S}$ and $\mathbf{S}_{\mathbf{k}}^{F}$, respectively. We note that this mapping is valid for any two-band model at half filling. The explicit form of $\mathbf{B}_{\mathbf{k}}(t)$ in the case of graphene subject to the circularly polarized electric field is given in the Supplementary Note 1.

For the parameters chosen [see Eq.~\eqref{eq:parameters}] these ground states have different topology: $\mathbf{S}_{\mathbf{k}}^{S}$ does not wrap around the Bloch sphere ($\text{Ch}=0$) while $\mathbf{S}_{\mathbf{k}}^{F}$ does ($\text{Ch}=-1$) (see Figs.~\ref{fig:infinite}{\bf{b}} and \ref{fig:infinite}{\bf{c}}). This implies that there is at least one $k$-point in the BZ for which the vectors $\mathbf{S}_{\mathbf{k}}^{S}$ and $\mathbf{S}_{\mathbf{k}}^{F}$ point in opposite directions (Supplementary Figure 1 and Supplementary Note 2) and, as a result, the overlap of the ground states is identically zero:
\begin{equation}
\left|\langle\psi_{0}^{S}|\psi_0^{F}\rangle\right|^{2}=
\prod_{\mathbf{k}}\,\left(\frac{1+\mathbf{S}_{\mathbf{k}}^{S}
\cdot\mathbf{S}_{\mathbf{k}}^{F}}{2}\right)=0
\label{eq:overlap}
\end{equation}

We can now consider the dynamical process by which the periodic driving is turned on. In principle, the Chern number inherits a time-dependence from the time-dependence of the pseudo spin configuration $\mathbf{S}(t)$ obtained by integrating the equation of motion \eqref{eq:equation-motion} subject to the initial condition 
$\mathbf{S}_{\mathbf{k}}=\mathbf{S}_{\mathbf{k}}^{S}$. However, a straightforward calculation shows that this is not the case. This follows from the fact that $\partial_{t}\text{Ch}$ can be written as:
\bea
\partial_{t}\text{Ch}(t)&=&\frac{1}{4\pi}\iint_{BZ}\, dk_{x}\, dk_{y}\,
\left\{ \left(\partial_{k_{y}}\mathbf{B}_{\mathbf{k}}(t)\cdot
\partial_{k_{x}}\mathbf{\mathbf{S}}_{\mathbf{k}}(t)\right)\right.
\nonumber\\ &&
-\left.\left(\partial_{k_{x}}\mathbf{\mathbf{B}}_{\mathbf{k}}(t)
\cdot\partial_{k_{y}}\mathbf{\mathbf{S}}_{\mathbf{k}}(t)\right)\right\}. 
\label{eq:intermediate}
\eea
If $\mathbf{S}_{\mathbf{k}}(t)$ and $\mathbf{B}_{\mathbf{k}}(t)$ are sufficiently smooth vector fields in the BZ then it follows that the expression above is identically zero (see Methods). From Eq.~\eqref{eq:equation-motion} one can see that an initially smooth pseudo spin configuration, $\mathbf{S}_{\mathbf{k}}(t)$ remains smooth under a smooth pseudo magnetic field $\mathbf{B}_{\mathbf{k}}(t)$. We can therefore formulate a no-go theorem as follows:
\begin{quote}
If the initial pseudo spin configuration is smooth (at least $\mathcal{\mathbb{C}}^{1}$) in the Brillouin
zone and the pseudo magnetic field is smooth (at least $\mathcal{\mathbb{C}}^{2}$), then the Chern number is conserved under the unitary evolution generated by the pseudo magnetic field.
\end{quote}
We note that: i) this theorem is valid for any two-band model at half filling for which the mapping in Eq.~\eqref{eq:mapping} applies, ii) the theorem holds even for time-dependent Hamiltonians and/or gapless Hamiltonians, as long as $\mathbf{B}_{\mathbf{k}}(t)$ is $\mathcal{\mathbb{C}}^{2}$ in the BZ for all times, and iii) smoothness in time is not required, i.e., our results also apply to sudden quenches for which the conservation of the Chern number has been noted before in various contexts~\cite{matt1,matt2,pedro}. We should stress that the smoothness of $\mathbf{B}_{\mathbf{k}}(t)$ in $\mathbf{k}$ is guaranteed by the locality of the $H(t)$ in real space, i.e., $\mathbf{B}_{\mathbf{k}}(t)$ can become singular in $\mathbf{k}$ only if the Hamiltonian $H(t)$ includes infinite range hopping in real space, and it is therefore not very restrictive. For example, the band structure of graphene is singular at the two Dirac points, but the pseudo magnetic field configuration:
\be
\begin{array}{c}
\mathbf{B}_{\mathbf{k}}=\left(1+\cos\left(\frac{3}{2}k_{x}-\frac{\sqrt{3}}{2}k_{y}\right)+
\cos\left(\frac{3}{2}k_{x}+\frac{\sqrt{3}}{2}k_{y}\right),\right.\\
\left.\sin\left(\frac{3}{2}k_{x}-\frac{\sqrt{3}}{2}k_{y}\right)+
\sin\left(\frac{3}{2}k_{x}+\frac{\sqrt{3}}{2}k_{y}\right),0\right)
\end{array}
\ee
is analytic in the BZ and satisfies the condition of the theorem.

The no-go theorem opens the question of whether it is experimentally possible to prepare a topologically nontrivial state by driving a topologically trivial one.

%%%%%%%%%%%%%%%%%%%%%%%%%%%%%%%%%%%%%%%%%%%%%%%%%%%%%%%%%%%%%%%%%%%%%%%
%%%%%%%%%%%%%%%%%%%%%%%%%%%%%%%%%%%%%%%%%%%%%%%%%%%%%%%%%%%%%%%%%%%%%%%

\subsection*{System with Boundaries}

\begin{figure*}[t!]
\includegraphics[width=1.95\columnwidth]{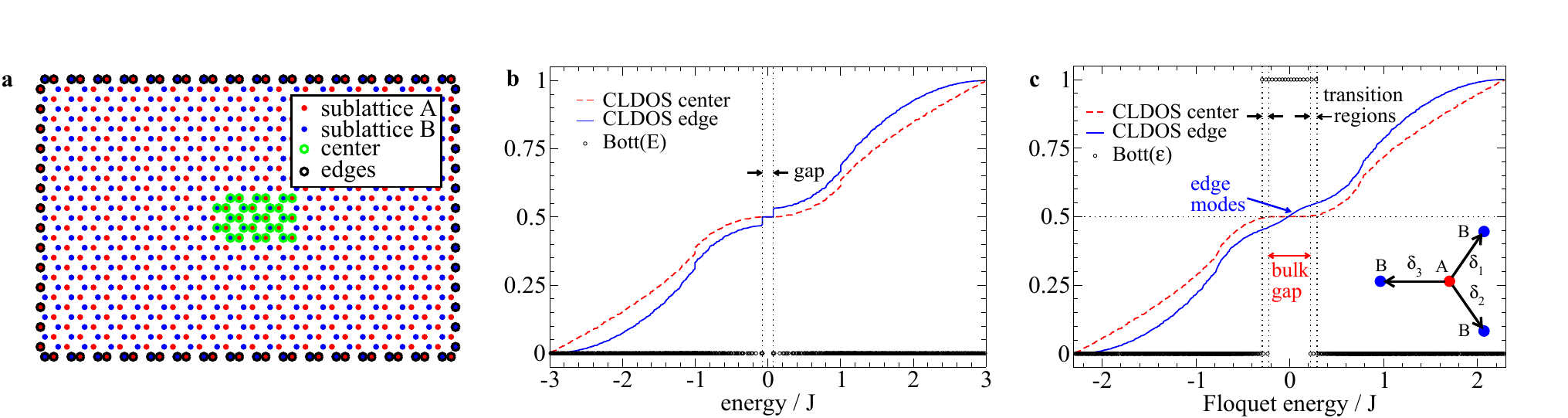}
\caption{\label{fig:boundary} Geometry and indicators for the system with boundaries. {\bf{a}}. One of the patch geometry considered. It contains a total of $928$ lattice sites evenly divided into the $\mathcal{A}$ (red dots) and $\mathcal{B}$ (blue dots) sublattices. The green and black circles indicate the sites defined as the center and the edges, respectively. These sites are used to compute the CLDOS. {\bf{b}} The CLDOS of ${H}_{S}$ is flat around $\varepsilon=0$ indicating a gap. The Bott index is identically zero for all energies. {\bf{c}}. For ${H}_{F}$, the CLDOS at the edges (at the center) has a finite (zero) slope about $\varepsilon=0$. This indicates the presence of edge states inside the bulk gap. Moreover, the Bott index for energies within the bulk gap is $+1$ indicating that the system is topological. Inset in {\bf{c}}. Site in sublattice $\mathcal{A}$ and its three nearest neighbors. The nearest neighbor vectors are: $\delta_{1,2}=(a/2)(1,\pm\sqrt{3})$, $\delta_3=a(-1,0)$.}
\end{figure*}

Experimental systems have boundaries, so here we address what happens when translational invariance is broken. We consider a finite, isolated system (such that the dynamics is unitary) with open boundary conditions (see the inset in Fig.~\ref{fig:boundary}{\bf{a}}). To characterize the topological properties of systems with broken translational symmetry one cannot rely on the Chern number. We use two complementary indicators: (i) the cumulative local density of states $\text{CLDOS}(\varepsilon)=\int_{-\infty}^{\varepsilon}d\varepsilon^\prime\ \text{LDOS}(\varepsilon^\prime)$ and (ii) the Bott index~\cite{bott}. 

The Bott index is a topological invariant that can be thought as the generalization of the Chern number for finite, non-translationally invariant, systems. Some remarkable properties of the Bott index are: i) it is computed directly in real space, ii) it is quantized for finite systems, and iii) it can be defined in a patch geometry. This is in contrast to the Chern number which: i) is computed by integrating the partial derivatives of the wave-function over a two-dimensional torus, ii) is generically non-quantized when the integration is replaced by a discrete sum and, iii) cannot be defined in a patch geometry. In Ref.~\cite{bott}, the Bott index was introduced for finite disordered two-dimensional systems with periodic boundary conditions (i.e., on a torus), but it can be straightforwardly generalized to other geometries (see Methods). In equilibrium, the Bott index is a function of the energy $\varepsilon$. It is computed by projecting special matrices (see Methods) onto the subspaces spanned by the eigenstates of the Hamiltonian with energies $\varepsilon^\prime<\varepsilon$. This definition assumes that the eigenstates with energy $\varepsilon^\prime<\varepsilon$ are fully occupied while the eigenstates with energies $\varepsilon^\prime>\varepsilon$ are empty. We have extended the Bott index definition to arbitrary sample geometries and non-equilibrium situations by taking into account the non-equilibrium character of the wave-function (see Methods). The numerical evidence gathered in this work strongly suggests that this generalized Bott index is a function of time and is quantized. However, this properties have not been proven rigorously.

In Fig.~\ref{fig:boundary}{\bf{a}}, we show one of the patch geometry considered and indicate the edge and bulk sites which have been used to compute the CLDOS. In Fig.~\ref{fig:boundary}{\bf{b}} and \ref{fig:boundary}{\bf{c}}, we show the CLDOS and the equilibrium Bott index for the static Hamiltonian ${H}_{S}$ and the Floquet Hamiltonian ${H}_{F}$, respectively. We stress that the Floquet Hamiltonian is computed exactly (see Methods). The $\text{CLDOS}(\varepsilon)$ of ${H}_{S}$ (both at the center of the sample and along the edges) has a plateau around $\varepsilon=0$ signifying that there are no states at $\varepsilon=0$, i.e., the system is gapped. Moreover, the Bott index is identically zero indicating that both ${H}_{S}$ and its ground state $\psi_0^{S}\rangle$  are topologically trivial. On the contrary, ${H}_{F}$ has edge states inside the bulk gap, as shown by the finite (zero) slope of the $\text{CLDOS}$ at the edge (center) for $\varepsilon\approx0$. The existence of topologically protected edge modes is confirmed by the Bott index. In fact, in equilibrium, the Bott index at some energy, $\text{Bott}(\varepsilon)$, is equal to the number of edge states at that energy. As one can see in Fig.~\ref{fig:boundary}{\bf{c}}, $\text{Bott}(\varepsilon\simeq0)=1$, indicating that the ground state of ${H}_{F}$ at half filling ($\varepsilon=0$) is topologically nontrivial. 

\begin{figure}[!t]
\includegraphics[width=0.9\columnwidth]{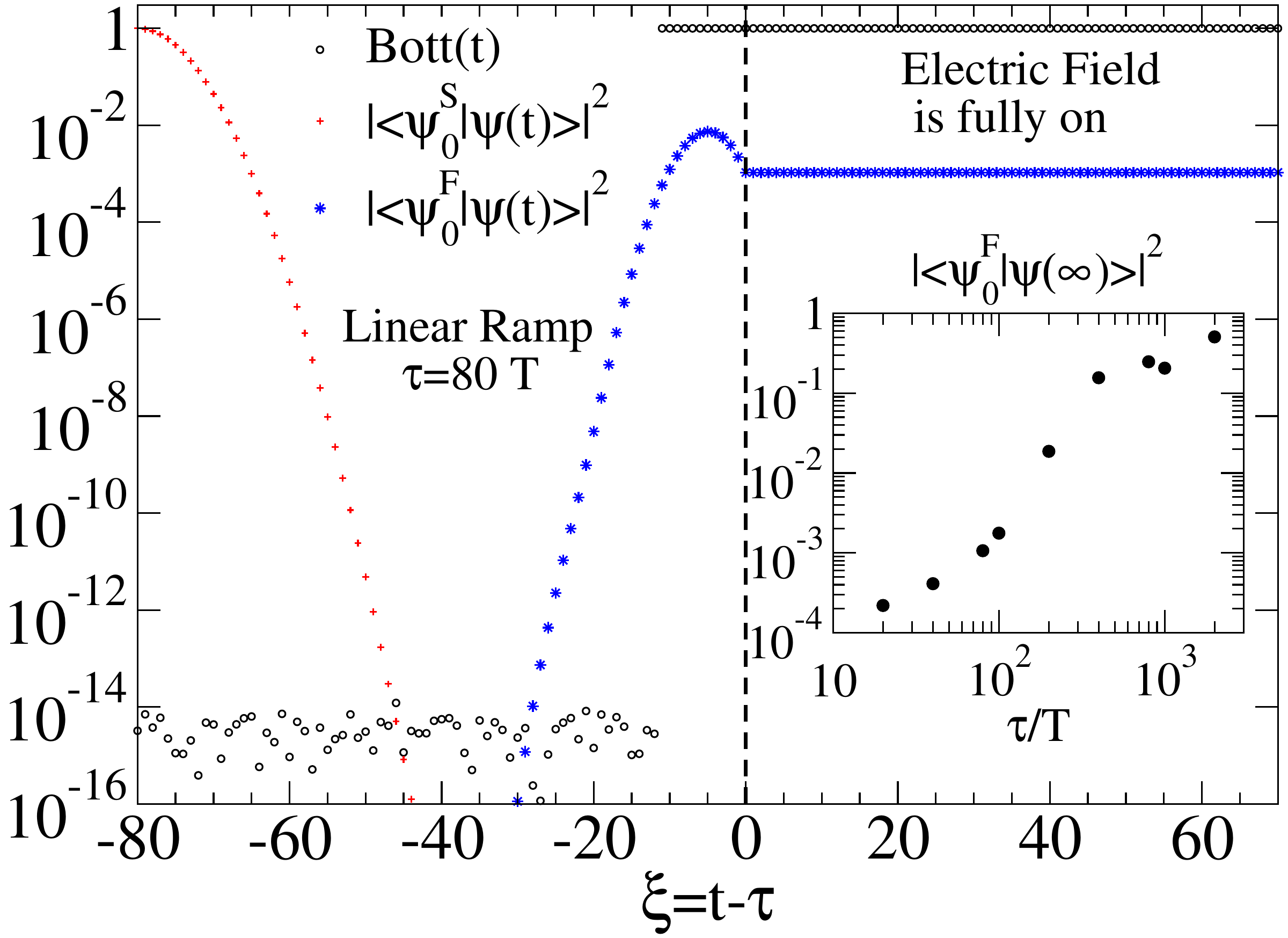}
\caption{\label{fig:ramp} Dynamical ramp with $\tau=80\, T$ and $N_\text{sites}=928$ in a system with boundaries. (Main) The evolution, at stroboscopic times $t_{n}=n\, T$, of the Bott index and the overlaps $\left|\langle\psi_{0}^{S}|\psi(t)\rangle\right|^{2}$ and $\left|\langle\psi_{0}^{F}|\psi(t)\rangle\right|^{2}$. For $\xi=t-\tau>0$, the electric field is fully on. The overlap $\left|\langle\psi_{0}^{F}|\psi(t_{n})\rangle\right|^{2}$ is a non-monotonic function of $t_{n}=n\, T$. (Inset) The overlap at then end of the dynamical ramp tends to increase with increasing $\tau$.}
\end{figure}

In order to study the adiabatic turning on of the periodic driving, we solve the time-dependent Schr\"odinger equation~\cite{marcos_evolution3,he_brown_14} subject to the initial condition $|\psi(t=0)\rangle=|\psi_0^{S}\rangle$ (see Methods). At stroboscopic times $t_{n}=n\, T$ during the time evolution, we monitor the Bott index and the overlaps of $|\psi(t)\rangle$ with both the initial state and the Floquet ground state:
\be
\text{Bott}(nT\,),\quad\left|\langle\psi_{0}^{S}|\psi(nT\,)\rangle\right|^{2},
\quad\left|\langle\psi_{0}^{F}|\psi(nT\,)\rangle\right|^{2}
\ee
Their behavior, for a system of size $N_\text{sites}=928$ and ramping time $\tau=80\,T$, are shown in Fig.~\ref{fig:ramp}. One can see that overlap with the initial state decays to zero rapidly while the overlap with the ground state of the Floquet Hamiltonian increases. For $t_{n}=n\, T>\tau$, the electric field has reached its final value and the overlap with $|\psi_0^{F}\rangle$ becomes independent on $n$ since $|\psi_0^{F}\rangle$ is an eigenstate of the evolution operator over a period:
\be
\left|\langle\psi_{0}^{F}|\psi(t_{n+1})\rangle\right|^{2}=
\left|\langle\psi_{0}^{F}|U(T\,)|\psi(t_{n})\rangle\right|^{2}=
\left|\langle\psi_{0}^{F}|\psi(t_{n})\rangle\right|^{2}.
\ee
Interestingly, for the parameters chosen, at $t\approx\tau-11T$ (i.e., slightly before the electric field is fully on) the Bott index jumps from zero and becomes one, indicating the wavefunction has acquired a topological character. We also note that the overlap with the Floquet ground state $|\psi_0^{F}\rangle$ increases non-monotonically with time. This suggests that the final overlap $\left|\langle\psi_{0}^{F}|\psi\left(t=\infty\right)\rangle\right|^{2}$ can be increased by using more sophisticated ramping protocols. For example, by instantaneously quenching the amplitude of the electric field from its value when the overlap $\left|\langle\psi_{0}^{F}|\psi\left(t\right)\rangle\right|^{2}$ has a local maximum to its final value. In the inset in Fig.~\ref{fig:ramp}, we plot the value of the overlap with the Floquet ground state at the end of the ramp, i.e., $\left|\langle\psi_{0}^{F}|\psi\left(t=\infty\right)\rangle\right|^{2}$, for different ramping times $\tau$ and observe that, as expected, it generally increases with increasing $\tau$. We note that the system can become topological even if the ramp is not adiabatic and, therefore, the overlap between $|\psi(t)\rangle$ and $|\psi_0^{F}\rangle$ is small.

\begin{figure*}
\includegraphics[width=1.95\columnwidth]{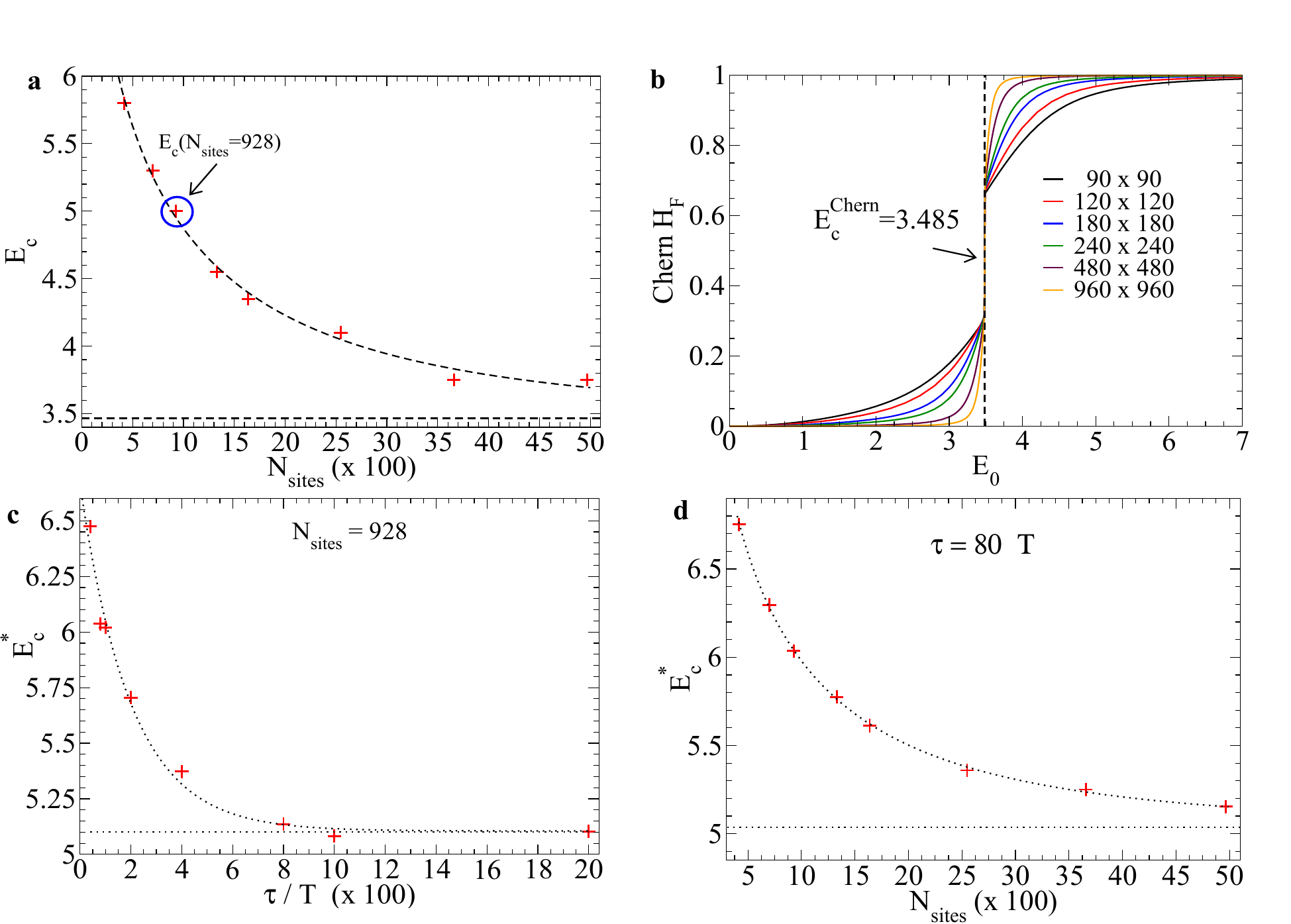}
\caption{\label{fig:EC(L,t)} Scaling analysis of the critical fields at which the systems become topological. {\bf{a}}. The equilibrium chern number of the translationally invariant system jumps discontinuously at $E_c^\text{chern}=3.485$ indicating that, for $E>E_c^\text{chern}$, the translationally invariant system becomes topological. {\bf{b}} The critical electric field $E_c$ at which the Floquet Hamiltonian becomes topological decreases exponentially with the linear size of the system, $L\approx \sqrt{N_\text{sites}}$, and approaches $E_c=3.47$ in the thermodynamic limit. This value is compatible with the value at which the translationally invariant system becomes topological (see panel {\bf{a}}). {\bf{c}}. For a fixed system size ($N_\text{sites}=928$), the value of the electric field $E_c^*$ at which the wavefunction becomes topological decreases exponentially with $\tau/T$ and approaches $E_c^*=5.10$ as $\tau/T\rightarrow\infty$. This value is compatible with the value at which $H_F$ becomes topological for $N_\text{sites}=928$ (see the point indicated by an arrow in panel {\bf{b}}). {\bf{d}} For fixed ramp time ($\tau=80 T$), critical field $E_c^*$ at which the wavefunction becomes topological decreases exponentially with $L$ and approaches $E_c^*=5.04$ in the thermodynamic limit.}
\end{figure*}

In order to relate the dynamical behavior of the non-equilibrium Bott index to the properties of $H_F$, we first study the critical field $E_c(N_\text{sites})$ at which the Floquet Hamiltonian becomes topological. For each system size, we compute the exact Floquet Hamiltonian for many different values of the electric field and repeat the analysis in Fig.~\ref{fig:boundary}{\bf{c}}. For weak electric fields, $\text{Bott}(\varepsilon)$ is identically zero, while for $E>E_c$ the Bott index is unity for some energies. This allows us to extract $E_c$ for different system sizes which is reported in Fig.~\ref{fig:EC(L,t)}{\bf{b}}. A fit to those results shows that $E_c$ approaches the infinite system size result exponentially in the linear dimension of the system ($L\approx \sqrt{N_\text{sites}}$). The infinite size result is, in turn, compatible with the value $E_c^\text{chern}=3.485$ at which the translationally invariant system becomes topological, as shown by the Chern number of the lowest Floquet band in Fig.~\ref{fig:EC(L,t)}{\bf{a}}.

\begin{figure*}[t!]
\includegraphics[width=1.95\columnwidth]{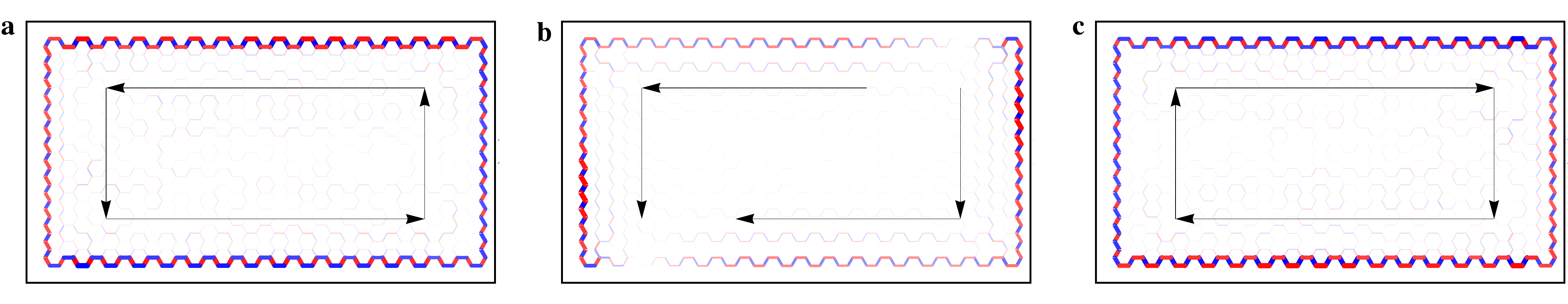}
\caption{\label{fig:non_eq_current} Time-averaged physical current after a dynamical ramp with $\tau=80$. We show results for systems with filling factors: $\nu=0.46,\,0.5,\,0.54$. The red color indicates a current flowing from sublattice $\mathcal{A}$ to $\mathcal{B}$ along the nearest neighbor vectors $\delta_i$: $\delta_{1,2}=(a/2)(1,\pm\sqrt{3})$, $\delta_3=a(-1,0)$. The blue color indicates the opposite direction. The black arrows indicate the direction of the edge current and are a guide to the eye. {\bf{a}}. For an initial filling $\nu=0.46$, the non-equilibrium current is concentrated along the edges and moves counter-clockwise. {\bf{c}}. For an initial filling $\nu=0.54$, the non-equilibrium current is concentrated along the edges and moves clockwise. {\bf{b}}. At half-filling, both chiralities are presents. On the top and left edge the current moves clockwise, while on the bottom and right edge it moves counter-clockwise. These edge currents are balanced by a current diffusing through the bulk. The non-equilibrium wave-functions corresponding to panels {\bf{a}} and {\bf{c}} are topologically trivial (Bott index is zero) while the one corresponding to panel {\bf{b}} is topological non-trivial (Bott index is one).}
\end{figure*}

Next, we study the value of the electric field at the time when the non-equilibrium Bott index jumps, i.e., $E_c^*(N_\text{sites})$. One could advance that, when the driving is turned on very slowly (i.e., adiabatically) $E_c^*(N_\text{sites})$ is identical to $E_c(N_\text{sites})$ for any given system size. This is indeed what we find. In Fig.~\ref{fig:EC(L,t)}{\bf{c}}, we show the critical field $E_c^*$ at which $|\psi(t)\rangle$ becomes topological, for a system with $928$ sites, as a function of the ramping time. A fit to the results shows that $E_c^*$ approaches the adiabatic limit (infinite time ramp) result exponentially in the ramp time. Our extrapolated result for the adiabatic limit is compatible with $E_c(N_\text{sites} = 928)\approx5.0$ for which $H_F$ becomes topological (see point signaled by an arrow in Fig.~\ref{fig:EC(L,t)}{\bf{b}}). We have also studied what happens if one fixes the ramp time and increases the system size, see Fig.~\ref{fig:EC(L,t)}{\bf{d}}. In this case, the critical field approaches the thermodynamic limit result also exponentially with the linear dimension of the system.

%%%%%% CURRENTS %%%%%%%%%%%

To make closer contact with experiments (such as Ref.~\cite{Esslinger}), we investigate the current that flows through the sample under driving (see Supplementary Figure 2 and Supplementary Note 3). This is, in principle, a measurable quantity~\cite{fregoso}. We stress that the physical current is different from the current one obtains using the Floquet Hamiltonian~\cite{marin1,marin2} (see Methods). The former connects only nearest neighbor sites, while the latter can flow between far away lattice sites if there are longer-range hopping terms in ${H}_F$. We compute the physical current by solving the time-dependent Schr\"odinger equation.

Contrary to the overlap, the average current over a period continues to evolve for $t>\tau$ (i.e., when the electric field amplitude has already reached its final value). Therefore, after the end of the ramp, we evolve the system for a large number of periods ($10^3$) so that the averaged current (over a period) becomes stationary. (We note that the instantaneous current still changes within a period.) Results for the average current are shown in Fig.~\ref{fig:non_eq_current}. Remarkably, we find that the chirality of the non-equilibrium current depends on the filling fraction (recall that the initial state is the ground state of $H_S$), see Supplementary Figure 3, 4, 5, 6 and 7 and Supplementary Note 4. To identify what changes when the Bott index jumps in systems at half filling, we have compared the currents for four different ramps (see Supplementary Figure 8 and 9, Supplementary Table 1, and Supplementary Note 5). The first three ramps (last ramp) are (is) such that the final value of the electric field is smaller (larger) than the critical value required for the Bott index to jump. For the last ramp, after the Bott index jumps to one, the currents are much larger than for the first three ramps and are localized along the edges of the system.

%%%%%%%%%%%%%%%%%%%%%%%%%%%%%%%%%%%%%%%%%%%%%%%%%%%%%%%%%%%%%%%%%%%%%%%%%%%%%%%%%%%%%%
%%%%%%%%%%%%%%%%%%%%%%%%%%%%%%%%%%%%%%%%%%%%%%%%%%%%%%%%%%%%%%%%%%%%%%%%%%%%%%%%%%%%%%

\section*{Discussion}

The two topological invariants studied here during the switching on of a periodic perturbation, the Chern number in translationally invariant (thermodynamically large) systems and the Bott index in systems with boundaries, exhibit qualitatively different behavior. The Chern number is conserved while the Bott index can change under unitary evolution. The conservation of the Chern number might appear surprising since, during any non-equilibrium process, some excitations are generated and the final state, which corresponds to a partially filled valence and conductance band, need not have a quantized Chern number. However, this argument does not take into account that, under unitary evolution, each quasi-momentum $\mathbf{k}$ is in a coherent superposition of the valence and conduction band. It is precisely this coherence that prevents the Chern number from changing. 

Our results for the Bott index show that, when one turns on a drive slowly starting from a topologically trivial state in finite systems with open boundary conditions, there is a critical field $E_c^*$ (which depends on the ramp time $\tau$) at which the Bott index jumps from zero to one. This indicates that the system becomes topologically nontrivial, even if the turn on is not adiabatic. If the drive is turned on adiabatically, $E_c^*$ approaches (with increasing system size) $E_c^\text{chern}=3.485$ at which the Chern number indicates that the Floquet Hamiltonian of the system with periodic boundary conditions becomes topological in the thermodynamic limit. This agrees with the intuition that an adiabatic turn on of the drive should allow one to generate a topologically nontrivial state, but is in stark contrast with the fact that the Chern number is a conserved quantity. Two possible explanations to these fundamental differences in thermodynamically large systems are: either dynamical topological transitions only occur in systems with boundaries or those transitions can happen in systems with and without boundaries, despite the fact that the Chern number does not change in the latter. Unfortunately, we cannot use the Bott index to discriminate between those possibilities because it is neither quantized nor a smooth function of the energy in clean system with periodic boundary conditions (see Supplementary Figure 10 and Supplementary Note 6). If there is a dynamical transition in clean system with periodic boundary conditions, another indicator needs to be found to identify it.

Closer to experiments, our results for the chiralities of the edge currents, namely, that they depend on the filling fraction, might also appear intriguing. They contrast with the fact, which we have checked, that the edge modes of a topological Floquet Hamiltonian support a single-particle current whose chirality is determined by the polarization of the electric field. Our results are a consequence of the fact that the current in many-particle non-equilibrium states has contributions from Floquet eigenstates with many different quasi-energies. While it is well known that a topological Floquet Hamiltonian supports chiral edge modes in the bulk gap, we have verified that other Floquet eigenstates can support currents with the opposite chirality (see Supplementary Figure 3, 4, 5, 6 and 7, and Supplementary Note 4). By changing the filling fraction one can change the contributions of different Floquet eigenstates and control the chirality of the current. This means that any potentially sharp signature of the topological transition (identified by the jump in the Bott index) in the many-particle current is smeared out by the contributions of Floquet eigenstates that are away from the bulk gap. The dependence of the chirality on the filling fraction is a strong prediction that can be tested in current experimental setups.

Our results open many new interesting questions: What is the nature of the topological transition in systems with boundaries? What is the dynamics of the edge states~\cite{edge-quench} across those transitions? What happens in the presence of interactions~\cite{marcos-interaction,luca-long-time} and/or a coupling to a bath? Which loading protocols maximize the occupation of the Floquet ground state? What are the distinctive signatures of non-topological wavefunctions evolving according to topological Hamiltonian (and vice versa)? Is the presence of deep lying current carrying Floquet eigenstates connected to the existence of new topological invariants unique to Floquet systems? Which physical observables capture best the time change of the Bott index? Is there a dynamical topological transition in clean system with periodic boundary conditions? If so, which topological indicator captures it? We hope our work will motivate further studies to address these and other related questions and to establish, in full mathematical rigor, the properties of the non-equilibrium Bott index introduced in this work.

%%%%%%%%%%%%%%%%%%%%%%%%%%%%%%%%%%%%%%%%%%%%%%%%%%%%%%%%%%%%%%%%%%%%%%%%%%%%%%%%%%%%%
%%%%%%%%%%%%%%%%%%%%%%%%%%%%%%%%%%%%%%%%%%%%%%%%%%%%%%%%%%%%%%%%%%%%%%%%%%%%%%%%%%%%%
%%%%%%%%%%%%%%%%%%%%%%%%%%%%%%%%%%%%%%%%%%%%%%%%%%%%%%%%%%%%%%%%%%%%%%%%%%%%%%%%%%%%%

\section*{Methods}

\subsection*{Chern number conservation under unitary evolution}

By going to momentum space, the system is parametrized as in Eq.~\eqref{eq:mapping}, where $\mathbf{S}_{\mathbf{k}}$ and $\mathbf{B}_{\mathbf{k}}$ are three dimensional time-dependent vectors fields. The Chern number of the occupied state is simply the number of wrapping of the pseudo-spin configuration around the Bloch sphere [see Eq. \eqref{eq:chern}]. The exact equation of motion is Eq.~\eqref{eq:equation-motion} (we have set $\hbar=1$). Putting together Eq.~\eqref{eq:mapping} and \eqref{eq:equation-motion}, we can perform the calculation using standard manipulations of classical vector fields. Our results are, however, fully quantum. We now compute $\partial_{t}\text{Ch}$: 

\bea
\partial_{t}\text{Ch}&=&\frac{1}{4\pi}\iint dk_{x}dk_{y}\,
\left\{ \dot{\mathbf{S}}\cdot\left[\partial_{x}\mathbf{S}\times\partial_{y}
\mathbf{S}\right]\right.\\&&+\left.\mathbf{S}\cdot\left[\partial_{x}\mathbf{\dot{S}}
\times\partial_{y}\mathbf{S}\right]+\mathbf{S}\cdot
\left[\partial_{x}\mathbf{S}\times\partial_{y}\mathbf{\dot{S}}\right]\right\},\nonumber 
\eea
where we have introduced the short hand notation $\partial_{k_{x}}\mathbf{S}=\partial_{x}\mathbf{S}=\mathbf{S}_{x}$, $\partial_{k_{y}}\mathbf{S}=\partial_{y}\mathbf{S}=\mathbf{S}_{y}$, and we have suppressed the suffix $\mathbf{k}$ in $\mathbf{S}_{\mathbf{k}}$. We consider the three terms separately:
\bea
I&=&\dot{\mathbf{S}}\cdot\left[\partial_{x}\mathbf{S}\times\partial_{y}
\mathbf{S}\right]=\left(\mathbf{S}\times\mathbf{B}\right)\cdot
\left(\partial_{x}\mathbf{S}\times\partial_{y}\mathbf{S}\right)\nonumber\\&=&
\left(\mathbf{S}\cdot\mathbf{S}_{x}\right)\left(\mathbf{B}\cdot
\mathbf{S}_{y}\right)-\left(\mathbf{S}\cdot\mathbf{S}_{y}\right)
\left(\mathbf{B}\cdot\mathbf{S}_{x}\right),
\eea
where we have used the evolution equation $\partial_{t}\mathbf{S}=\mathbf{S}\times\mathbf{B}$
and the Binet-Cauchy identity:
\be
\left(\mathbf{a}\times\mathbf{b}\right)\cdot\left(\mathbf{c}\times\mathbf{d}\right)=
\left(\mathbf{a}\cdot\mathbf{c}\right)\left(\mathbf{b}\cdot\mathbf{d}\right)-
\left(\mathbf{a}\cdot\mathbf{d}\right)\left(\mathbf{b}\cdot\mathbf{c}\right).
\ee
The second and third term are more involved. For example:
\bea
&&II=\mathbf{S}\cdot\left[\partial_{x}\mathbf{\dot{S}}\times\partial_{y}
\mathbf{S}\right]=\partial_{x}\mathbf{\dot{S}}\cdot
\left[\partial_{y}\mathbf{S}\times\mathbf{S}\right]\\&&=
\partial_{x}\left(\mathbf{S}\times\mathbf{B}\right)\cdot
\left[\partial_{y}\mathbf{S}\times\mathbf{S}\right]=
\left(\mathbf{S}_{x}\times\mathbf{B}+\mathbf{S}\times\mathbf{B}_{x}\right)
\cdot\left[\mathbf{S}_{y}\times\mathbf{S}\right],\nonumber
\eea
where we have used the distribution property of the cross product $\partial_{x}\left(\mathbf{S}\times\mathbf{B}\right)=\mathbf{S}_{x}\times\mathbf{B}+\mathbf{S}\times\mathbf{B}_{x}$. One can apply the Binet-Cauchy identity to obtain:
\bea
II&=&\left(\mathbf{S}_{x}\cdot\mathbf{S}_{y}\right)
\left(\mathbf{B}\cdot\mathbf{S}\right)-\left(\mathbf{S}_{x}
\cdot\mathbf{S}\right)\left(\mathbf{B}\cdot\mathbf{S}_{y}\right)\nonumber\\&&+
\left(\mathbf{S}\cdot\mathbf{S}_{y}\right)\left(\mathbf{B}_{x}
\cdot\mathbf{S}\right)-\left(\mathbf{S}\cdot\mathbf{S}\right)
\left(\mathbf{B}_{x}\cdot\mathbf{S}_{y}\right).
\eea
In a similar way, we get:
\bea
III&=&-\left(\mathbf{S}_{x}\cdot\mathbf{S}_{y}\right)\left(\mathbf{B}
\cdot\mathbf{S}\right)+\left(\mathbf{S}_{y}\cdot\mathbf{S}\right)
\left(\mathbf{B}\cdot\mathbf{S}_{x}\right)\nonumber\\&&-\left(\mathbf{S}
\cdot\mathbf{S}_{x}\right)\left(\mathbf{B}_{y}\cdot\mathbf{S}\right)+
\left(\mathbf{S}\cdot\mathbf{S}\right)\left(\mathbf{B}_{y}\cdot\mathbf{S}_{x}\right).
\eea
Putting all together, and carrying out the cancellations, we get:
\bea
\partial_{t}\text{Ch}&=&\frac{1}{4\pi}\iint\,\left(I+II+III\right)\nonumber\\&=&
\frac{1}{4\pi}\iint\,\left\{ \left(\mathbf{S}\cdot\mathbf{S}_{y}\right)
\left(\mathbf{B}_{x}\cdot\mathbf{S}\right)-\left(\mathbf{S}\cdot
\mathbf{S}\right)\left(\mathbf{B}_{x}\cdot\mathbf{S}_{y}\right)\right.\nonumber\\&&\left.\quad+
\left(\mathbf{S}\cdot\mathbf{S}\right)\left(\mathbf{B}_{y}\cdot
\mathbf{S}_{x}\right)-\left(\mathbf{S}\cdot\mathbf{S}_{x}\right)
\left(\mathbf{B}_{y}\cdot\mathbf{S}\right)\right\}.
\eea
So far this expression is completely general. Now we use that the initial state is a pure state:
\be
1=\text{Tr}\left[\rho^{2}\right]=\text{Tr}\left[\left(\frac{1_{2\times2}+
\mathbf{S}\cdot\mathbf{\mathbf{\sigma}}}{2}\right)^{2}\right]=
\frac{1+\mathbf{S}\cdot\mathbf{S}}{2},
\ee
from which it follows that $\mathbf{S}\cdot\mathbf{S}=1$, i.e., $\mathbf{S}$ is a unit vector for any point in the BZ. We then observe that:
\be
\left(\mathbf{S}\cdot\mathbf{S}_{y}\right)=\frac{1}{2}\partial_{y}
\left(\mathbf{S}\cdot\mathbf{S}\right)=0,\quad\left(\mathbf{S}\cdot
\mathbf{S}_{x}\right)=\frac{1}{2}\partial_{x}\left(\mathbf{S}\cdot\mathbf{S}\right)=0.
\ee
Therefore we arrive at Eq.~\eqref{eq:intermediate} in the main text. We observe that: 
\bea
\left(\mathbf{B}_{y}\cdot\mathbf{S}_{x}\right)&=&\partial_{x}\left(\mathbf{B}_{y}
\cdot\mathbf{S}\right)-\left(\mathbf{B}_{x,y}\cdot\mathbf{S}\right),\nonumber\\
\quad\left(\mathbf{B}_{x}\cdot\mathbf{S}_{y}\right)&=&\partial_{y}
\left(\mathbf{B}_{x}\cdot\mathbf{S}\right)-\left(\mathbf{B}_{y,x}\cdot\mathbf{S}\right).
\eea
If the vector field $\mathbf{B}$ is smooth the mixed derivative
commute, i.e., $\mathbf{B}_{x,y}=\mathbf{B}_{y,x}$, and we arrive
at:
\be
\partial_{t}\text{Ch}=\frac{1}{4\pi}\iint\, dk_{x}dk_{y}\left\{ 
\partial_{x}\left(\mathbf{B}_{y}\cdot\mathbf{S}\right)-
\partial_{y}\left(\mathbf{B}_{x}\cdot\mathbf{S}\right)\right\}.
\ee
If $\mbox{\ensuremath{\partial_{x}\left(\mathbf{B}_{y}\cdot\mathbf{S}\right)}}$
and $\partial_{y}\left(\mathbf{B}_{x}\cdot\mathbf{S}\right)$ are
continuous, we can use the periodicity of $\mathbf{B}$ and $\mathbf{S}$
in the Brillouin zone to obtain:
\be
\partial_{t}\text{Ch}=0.
\ee
We note that, up to this point, we have simply shown that $\dot{\text{Ch}}(t)=0$ if $\mathbf{S}(t)$ and $\mathbf{B}(t)$ are sufficiently smooth and $\mathbf{S}(t)$ represents a pure state, i.e., $\mathbf{S}\cdot\mathbf{S}=1$. However, to prove that the Chern number is conserved at all times, we still need to prove that, under time evolution, (i) a pure state remains pure and (ii) a smooth pseudo-spin configuration remains smooth. To verify that this is indeed the case, we look into the equation of motion \eqref{eq:equation-motion}. We note that, under this equation, the length of the vector $\mathbf{S}$ is conserved, i.e., $\partial_{t}\left(\mathbf{S}\cdot\mathbf{S}\right)=0$, and therefore the condition (i) is verified. We also note that the equation of motion is a linear differential equation. If $\mathbf{S}(t=0)$ is smooth in $k_{x},k_{y}$ and $\mathbf{B}(t)$ is smooth in $k_{x},k_{y}$ then $\mathbf{S}(t)$ remains smooth at all times. Therefore, the condition (ii) is verified. This concludes the proof of the theorem.

The statement that the Chern number cannot change independently of the time-evolution considered is similar to the result that, under unitary evolution, the von Neumann entropy is conserved. Both results do not predict the exact wavefunction at the end of a dynamical process but constrain the possible outcomes. 

\subsection*{\label{sub:Appendix.-Bott index}Bott index for out-of-equilibrium systems}

Consider a single-particle Hamiltonian (defined by a matrix $H$) on a lattice. Given the two diagonal matrices $X_{i,j}=x_{i}\,\delta_{i,j}$ and $Y_{i,j}=y_{i}\,\delta_{i,j}$, where $x_{i}$ and $y_{i}$ are the coordinates of the $i^{th}$ lattice site, we defined two unitary matrices:
\be
(U_{x})_{i,j}=\exp\left[i2\pi\frac{X_{i,j}}{L_{x}}\right],\quad 
(U_{y})_{i,j}=\exp\left[i2\pi\frac{Y_{i,j}}{L_{y}}\right]
\ee
where $L_{x,y}$ are the linear dimensions of the system. Let $R$ be the projector onto the eigenstates with up to energy $\varepsilon$, i.e., $R\equiv\sum_{\varepsilon^\prime<\varepsilon}|\varepsilon\rangle\langle\varepsilon|$. In equilibrium, the Bott index at energy $\varepsilon$ is defined as~\cite{bott}:
\be
\text{Bott}(\varepsilon)=\frac{1}{2\pi}\text{Im}\left[\text{Tr}
\left(\log\left(\tilde{U}_{y}\tilde{U}_{x}
\tilde{U}_{y}^{\dagger}\tilde{U}_{x}^{\dagger}\right)\right)\right],
\label{eq:definition-bott}
\ee
where $\tilde{U}_{x}=R\, U_{x}\, R$ and $\tilde{U}_{y}=R\, U_{y}\, R$ are the matrices $U_{x}$ and $U_{y}$ projected onto the states with up to energy $\varepsilon$.

In Ref.~\cite{bott}, the Bott index was defined on a torus geometry (i.e., for $H$ with periodic boundary conditions). We generalize the Bott index to other geometries and non-equilibrium situations by properly modifying the projector $R$. For example, one can change the boundary conditions in $H$ from periodic to open by switching off some hopping elements. Using the projector constructed with the eigenstates of $H$ with open boundary conditions in Eq.~\eqref{eq:definition-bott}, one can compute the Bott index in a patch geometry. We further generalize the Bott index to non-equilibrium situations by taking into account the occupation of the states during the dynamics. The Bott index of the occupied states is obtained by replacing the matrices $\tilde{U}_{x}$ and $\tilde{U}_{y}$ with the matrices $U_{x}(t)$ and $U_{y}(t)$:
\be
U_{x}(t)=R(t)\, U_{x}\, R(t)\quad U_{y}(t)=R(t)\, U_{y}\, R(t)
\ee
where $R(t)$ is the projector onto states occupied at time $t$. For example, if one has $R=|\alpha\rangle\langle\alpha|+|\beta\rangle\langle\beta|$ at $t=0$, at time $t$ the projector becomes $R(t)=|\alpha(t)\rangle\langle\alpha(t)|+|\beta(t)\rangle\langle\beta(t)|$, where $|\alpha(t)\rangle=U(t)|\alpha\rangle$ and $|\beta(t)\rangle=U(t)|\beta\rangle$ are the time-evolved states.

One expects the generalized Bott index to be quantized if $R$ is a sufficiently local projector \cite{bott}, i.e., $R_{i,j}$ is small if sites $i$ and $j$ are far from each other. Strong numerical evidence supporting the expectation that the non-equilibrium Bott index defined on the patch geometry is quantized is provided in the main text. However, we stress that a mathematically rigorous proof is lacking at the moment. On the other hand, the Bott index in clean systems (the systems in Ref.~\cite{bott} had disorder) on a torus is neither quantized nor a smooth function of the energy (see Supplementary Note 6). We expect this to be because the eigenstates of $H$ for this problem are plane waves, which are non-local in real space.

\subsection*{\label{sub:Appendix.-currents} Currents}

To identify the current operator, we look into the time-derivative of the site occupations:
\begin{equation}
n_{l}=c_{l}^{\dagger}c_{l}\rightarrow\frac{\partial n_{l}}{\partial t}=
\frac{\partial c_{l}^{\dagger}}{\partial t}c_{l}+c_{l}^{\dagger}
\frac{\partial c_{l}}{\partial t}\label{eq:dn_dt}.
\end{equation}
The equation of motion of $c_{l}$ is $\frac{\partial c_{l}}{\partial t}=i\left[H,c_{l}\right]=-i\sum_{m}H_{l,m}c_{m}$, where we have used the fact that any noninteracting fermionic Hamiltonian has the form $H=\sum_{n,m}H_{n,m}c_{n}^{\dagger}c_{m}$. Similarly, we compute $\frac{\partial c_{l}^{\dagger}}{\partial t}=i\sum_{m}H_{m,l}c_{m}^{\dagger}$. Substituting these expressions in Eq.~\eqref{eq:dn_dt}, and computing the expectation value, one obtains:
\begin{eqnarray}
\left\langle\frac{\partial n_{l}}{\partial t}\right\rangle&=&i\sum_{m}H_{m,l}\,\langle 
c_{m}^{\dagger}c_{l}\rangle-i\sum_{m}H_{l,m}\,\langle c_{l}^{\dagger}c_{m}\rangle
\nonumber\\&=&
\sum_{m\ne l}\,\mathcal{I}\left[2H_{l,m}\langle c_{l}^{\dagger}c_{m}\rangle\right]
\end{eqnarray}
where $\mathcal{I}\left[.\right]$ indicates the imaginary part, and we have used that $H_{m,l}=\overline{H_{l,m}}$ and $\langle c_{m}^{\dagger}c_{l}\rangle=\overline{\langle c_{l}^{\dagger}c_{m}\rangle}$ and the overline indicates complex conjugation. The continuity equation relates the time-derivative of the local density to the net current: $\langle\frac{\partial n_{l}}{\partial t}\rangle=\sum_{m\ne l}\, J_{m\rightarrow l}$. This allows us to identify the current flowing from site $m$ to site $l$ as:
\be
J_{m\rightarrow l}=\mathcal{I}\left[2H_{l,m}
\mbox{\ensuremath{\langle c_{l}^{\dagger}c_{m}\rangle}}\right]
\ee
It is crucial that the Hamiltonian that appears in the equation of motion, and in the definition of the current, is the time-dependent Hamiltonian $H(t)$ and not the Floquet Hamiltonian $H_{F}$. In general, the matrix elements of $H(t)$ and $H_{F}$ are different. Hence, the current computed using the Floquet Hamiltonian is, in general, not equal to the current that will be measured in experiments \cite{marin1,marin2}.

For graphene subjected to circularly polarized electric field, $H(t)$ contains only nearest neighbor terms. This leads to a current flowing only between nearest-neighbor lattice sites, while $H_{F}$ contains longer range hopping, which lead to a current flowing between distant sites. In general, the exact current $J_{m\rightarrow l}$ is time-dependent because both the matrix element $H_{l,m}$ and the expectation value $\langle c_{l}^{\dagger}c_{m}\rangle$ are time-dependent. Moreover, in non-equilibrium situations, the site occupancies are non-stationary $\langle\frac{\partial n_{l}}{\partial t}\rangle\ne0$. This implies that the instantaneous current is not locally conserved, i.e., $\sum_{m}\, J_{m\rightarrow l}\ne0$. We have averaged the instantaneous current over a full driving period in order to obtain $J^\text{ave}_{m\rightarrow l}$, which is approximately conserved, i.e., $\sum_{m}\, J^\text{ave}_{m\rightarrow l}\approx0$.

\subsection*{\label{sub:Appendix.-Heff} Numerical simulations for system with boundaries}
The time-dependent Hamiltonian is given by Eq.~\eqref{eq:def_H(t)}. Because of its noninteracting character, this problem can be efficiently solved in the single-particle basis \cite{marcos_evolution3,he_brown_14}. We denote as $H_{S}$ and $H_{1}(t)$ the $N_\text{sites}\times N_\text{sites}$ matrices ($N_\text{sites}$ being the number of lattice sites) that represent the static and time-dependent parts of the Hamiltonian in real space.

The evolution operator over a cycle is given by:
\be
U(T,0)=\prod_{j=0}^{N-1}U(t_{j+1},t_{j})
\ee
where $t_{j}=j\,\frac{T}{N}$ and $U(t_{j+1},t_{j})$ is computed using a second-order Trotter-Suzuki decomposition~\cite{trotter,suzuki,suzuki-trotter}:
\be
U(t+\delta t,t)=e^{-\frac{i\delta t}{2}\, H_{1}\left(t+\frac{\delta t}{2}\right)}\ 
e^{-i\delta t\, H_{S}}\ e^{-\frac{i\delta t}{2}\, 
H_{1}\left(t+\frac{\delta t}{2}\right)}\label{eq:U}.
\ee
Since $H_{S}$ is time-independent, $e^{-i\delta t\, H_{S}}$ needs to be computed only once (this is done by diagonalizing $H_{S}$). This leaves the computation of $e^{-\frac{i\delta t}{2}\, H_{1}\left(t+\delta t/2\right)}$, from the already diagonal $H_{1}(t)$, to be computed at each time step. By exact diagonalization of $U(T,0)$, we obtain the Floquet eigenstates and eigenvalues:
\be
U(T,0)=\sum_{l}|\psi_{l}\rangle e^{-i\theta_{l}}\langle\psi_{l}|=
\sum_{l}|\psi_{l}\rangle e^{-\frac{i}{\hbar}\,\varepsilon_{l}T}\langle\psi_{l}|
\ee
from which the single-particle Floquet Hamiltonian can be explicitly built as $H_{F}=\sum_{l}|\psi_{l}\rangle\varepsilon_{l}\langle\psi_{l}|$ where $\varepsilon_{l}=\frac{\hbar}{T}\theta_{l}$. We note that this procedure is not limited to high frequency driving and gives the numerically exact Floquet Hamiltonian $H_F$. The time-discretization $\delta t$ is chosen small enough to ensure that it does not affect the results. The lowest energy single-particle eigenstates of $H_{F}$ ($H_S$) are then collected into a rectangular matrix $W_F$ ($W_S$) of size $N_\text{sites}\times N_p$, where $N_p$ is the number of particles in the ground state (at half filling $N_p=N_\text{sites}/2$). For the parameters chosen [see Eq.~\eqref{eq:parameters}] the Floquet phases $\theta_{l}$ do not span the entire range $\left[-\pi,\pi\right]$ and therefore an unambiguous separation of the states in the ``Floquet valence band'' and ``Floquet conduction band'' is possible.

The time evolution of the many-particle system is obtained by multiplying the matrix $W_S$ from the left with the square matrix $U(T,0)$ of size $N_\text{sites}\times N_\text{sites}$. The overlaps between many-particle wavefunctions can also be easily computed as determinants of products of matrices such as $W_F$ and $W_S$, and their adjoints \cite{marcos_evolution3,he_brown_14}. Moreover, the $\langle c_{i}^{\dagger}c_{j}\rangle$ elements of the equal-time single-particle density matrix are given by the $i,j$ element of the square matrix $\overline{W_0 W_0^{\dagger}}$ of size $N_\text{sites}\times N_\text{sites}$.

\subsection*{\label{sub:cont_time} Non-stroboscopic times}

We have also computed wavefunctions overlaps and the Bott index at non-stroboscopic times. The overlap between the time-evolving state and the Floquet Fermi sea does not change after the electric field is fully on, with the Floquet Hamiltonian computed from $U(t+T,t)=\exp\left[-i H_F[t]T\right]$. (Note that the Floquet Hamiltonian depends on the choice of the initial time of period. However Floquet Hamiltonian corresponding to different choices of the initial time are unitary equivalent to each other, see for example ~\cite{marin1}.) We also find that, for all $t\ge t_n^*$, the Bott index does not change with time, i.e., $\text{Bott}(t\ge t_n^*)=1$. Here $t_n^*$ is the first stroboscopic times at which the Bott index becomes unity. However, we found that just before the transition, in our case for times $t\in\left(t_n^*-2 T,\,t_n^*\right)$, the Bott index at non-stroboscopic times can jump back and forth between zero and one. This is similar, and probably related, to the oscillations observed in the (equilibrium) Bott index as the Fermi energy enters in the bulk gap [see Fig.~\ref{fig:boundary}{\bf{c}}].

%\bibliographystyle{ieeetr}
%\bibliographystyle{ieeetr}
%\bibliography{NoGo}
%\bibliography{NoGo}

\section*{Acknowledgments}
We thank M. Bukov, T. Iadecola, D. Iyer, M. Kolodrubetz, T. C. Lang, A. Polkovnikov, 
and K. Sun for illuminating discussions. We thank T. C. Lang for sharing his software used to plot 
currents. This work was supported by the Office of Naval Research.

\section*{Author contributions}
L.D. performed the numerical simulations. L.D and M.R. wrote the paper.

\section*{Competing financial interests}
The authors declare no competing financial interests.

%%%%%%%%%%%%%%%%%%%%%%%%%%%%%%%%%%%%%%%%%%%%%%%%%%%%%%%%%%%%%%%%%%%%%%%%%%%%%%%%%%%
%%%%%%%%%%%%%%%%%%%%%%%%%%%%%%%%%%%%%%%%%%%%%%%%%%%%%%%%%%%%%%%%%%%%%%%%%%%%%%%%%%%
%%%%%%%%%%%%%%%%%%%%%%%%%%%%%%%%%%%%%%%%%%%%%%%%%%%%%%%%%%%%%%%%%%%%%%%%%%%%%%%%%%%
\newpage
\newpage
\begin{widetext}
\newpage
\setcounter{page}{1}
\setcounter{figure}{0}
\pagestyle{empty}
\renewcommand{\figurename}{Supplementary Figure}
\renewcommand{\tablename}{Supplementary Table}

%\section*{Supplementary Figures}

\begin{minipage}{\linewidth}% to keep image and caption on one page
\makebox[\linewidth]{%   
\includegraphics[width=0.37\columnwidth]{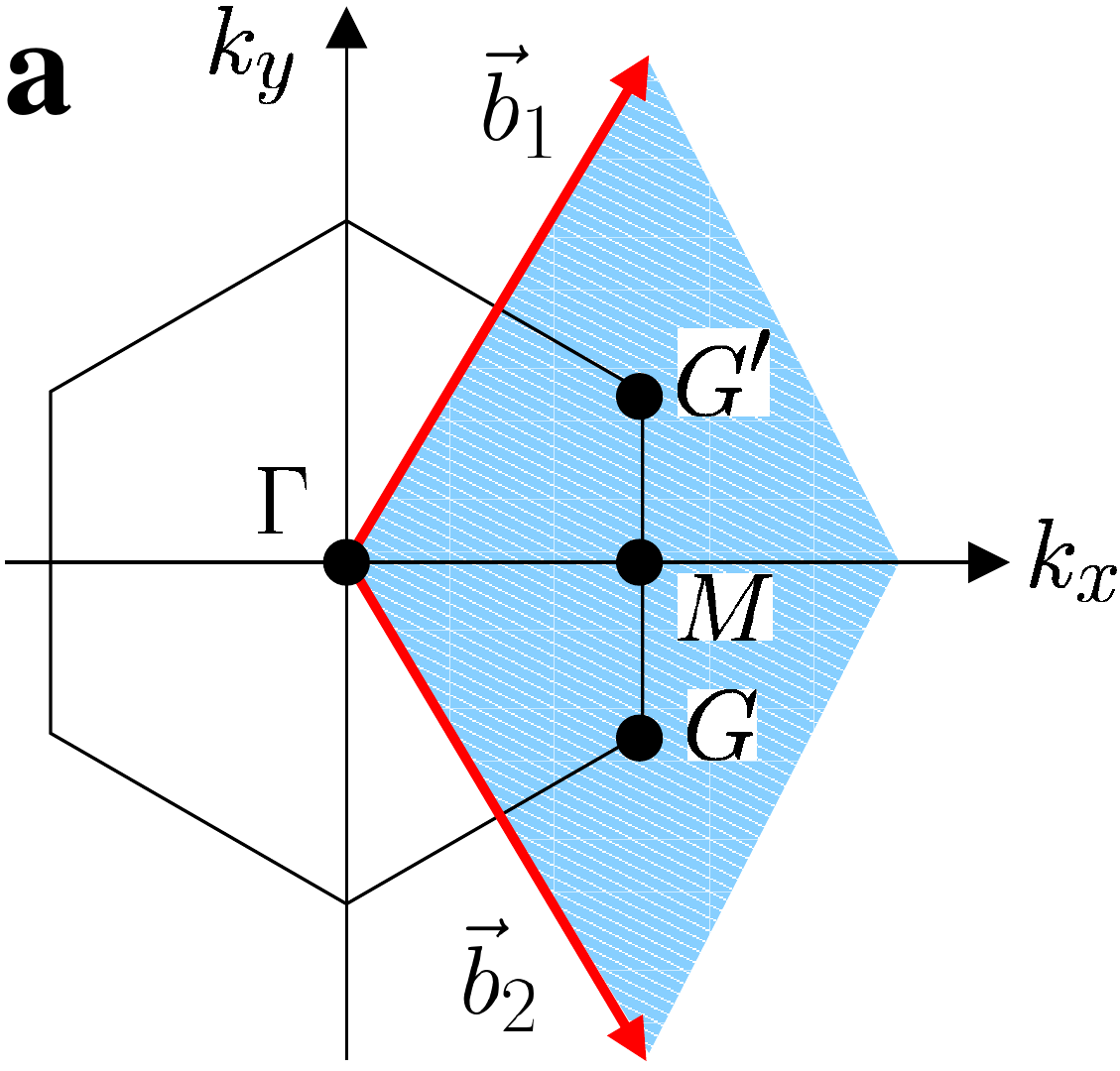}\,\quad 
\includegraphics[width=0.37\columnwidth]{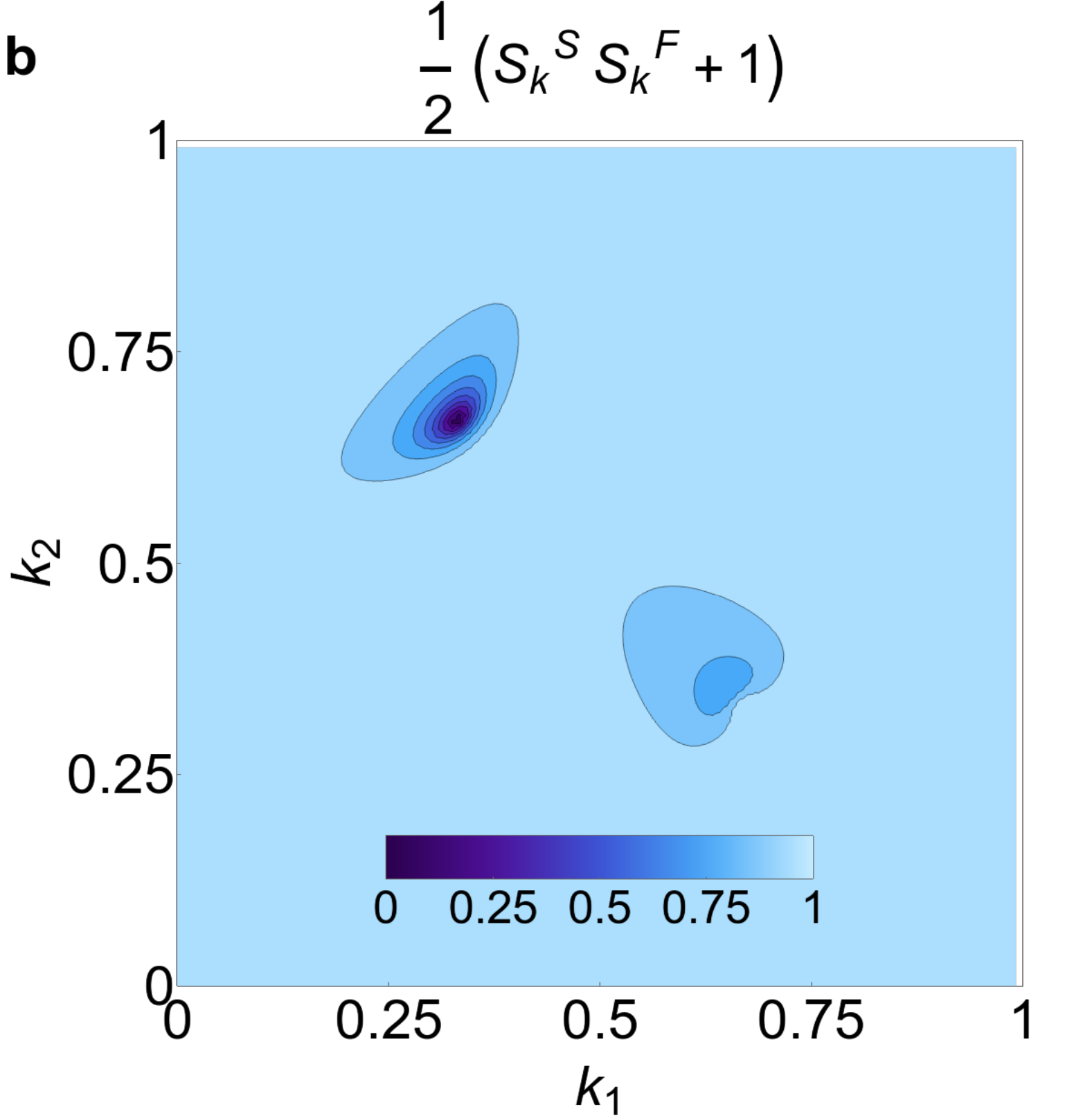}}
\captionof{figure}{\label{fig:BZ} The infinite translationally invariant system.
{\bf{a}}. The Brillouin zone (BZ) of graphene (shaded blue region) and its
high symmetry points $\Gamma,G,G^{\prime},M$. The reciprocal-lattice
vectors are $\vec{b}_{1,2}=\frac{2\pi}{3a}\left(1,\pm\sqrt{3}\right)$.
The BZ corresponds to $\vec{k}=k_{1}\vec{b}_{1}+k_{2}\vec{b}_{2}$
where $k_{1,2}\in[0,1)$. {\bf{b}}. The momentum resolved overlap between $|\psi_0^{S}\rangle$
and $|\psi_0^{F}\rangle$ is encoded in $\left(1+
\mathbf{S}_{\mathbf{k}}^{S}\cdot\mathbf{S}_{\mathbf{k}}^{F}\right)/2$ (see Eq.~\eqref{eq:overlap} in the main text). The pseudo spin configurations $\mathbf{S}_{\mathbf{k}}^{S}$
and $\mathbf{S}_{\mathbf{k}}^{F}$ differs significantly only
around the Dirac points.} 
\end{minipage}

\begin{minipage}{\linewidth}% to keep image and caption on one page
\makebox[\linewidth]{%  
\includegraphics[width=0.30\columnwidth]{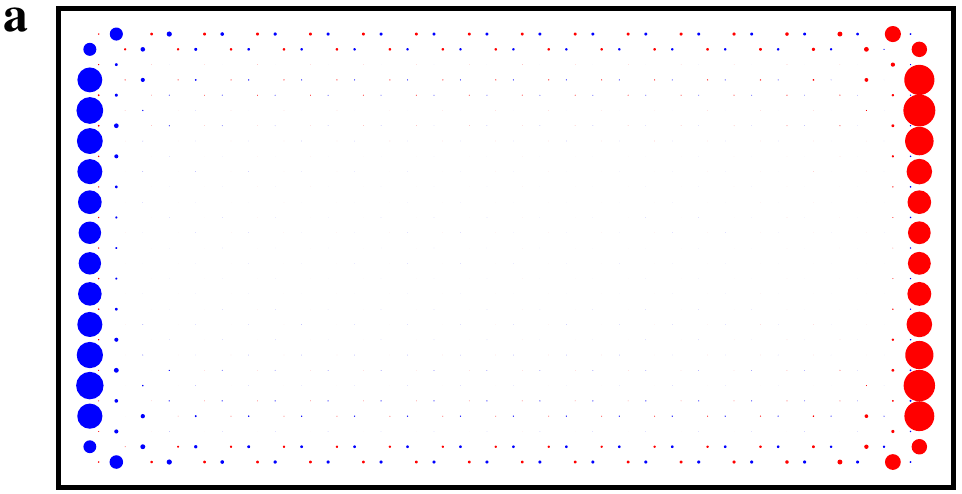}
\includegraphics[width=0.30\columnwidth]{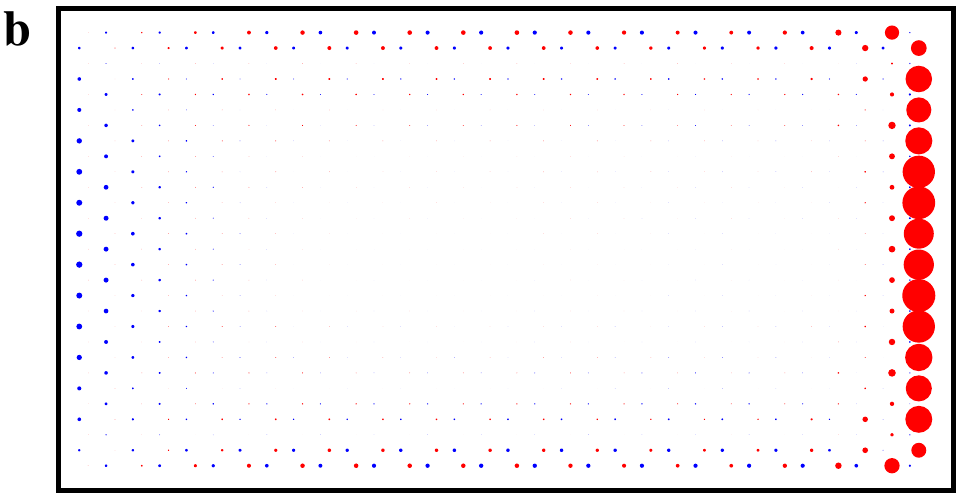}
\includegraphics[width=0.30\columnwidth]{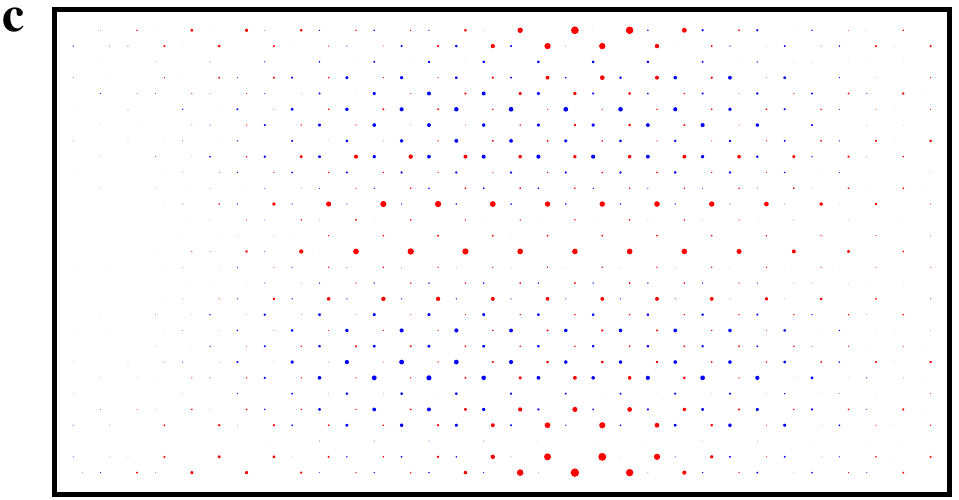}}
\captionof{figure}{ Single-particle Floquet eigenstates. The
size of the disks represents the occupation of a specific lattice site.
The red and blue disks indicate lattice sites that belong to the $\mathcal{A}$
and $\mathcal{B}$ sublattices, respectively. Panels {\bf{a}}, {\bf{b}}, and {\bf{c}} 
correspond to eigenstates with energies $\varepsilon/J\approx-0.01,\ -0.18,\ -0.31$,
respectively. The eigenstates in panels {\bf{a}} and {\bf{b}} are localized
on the edges and their energies lie in the bulk gap. The wavefunction
in panel {\bf{c}} is localized in the bulk and its energy lies outside the gap.\label{fig:edges-wf}}
\end{minipage}

\begin{minipage}{\linewidth}% to keep image and caption on one page
\makebox[\linewidth]{%  
\includegraphics[width=0.45\columnwidth]{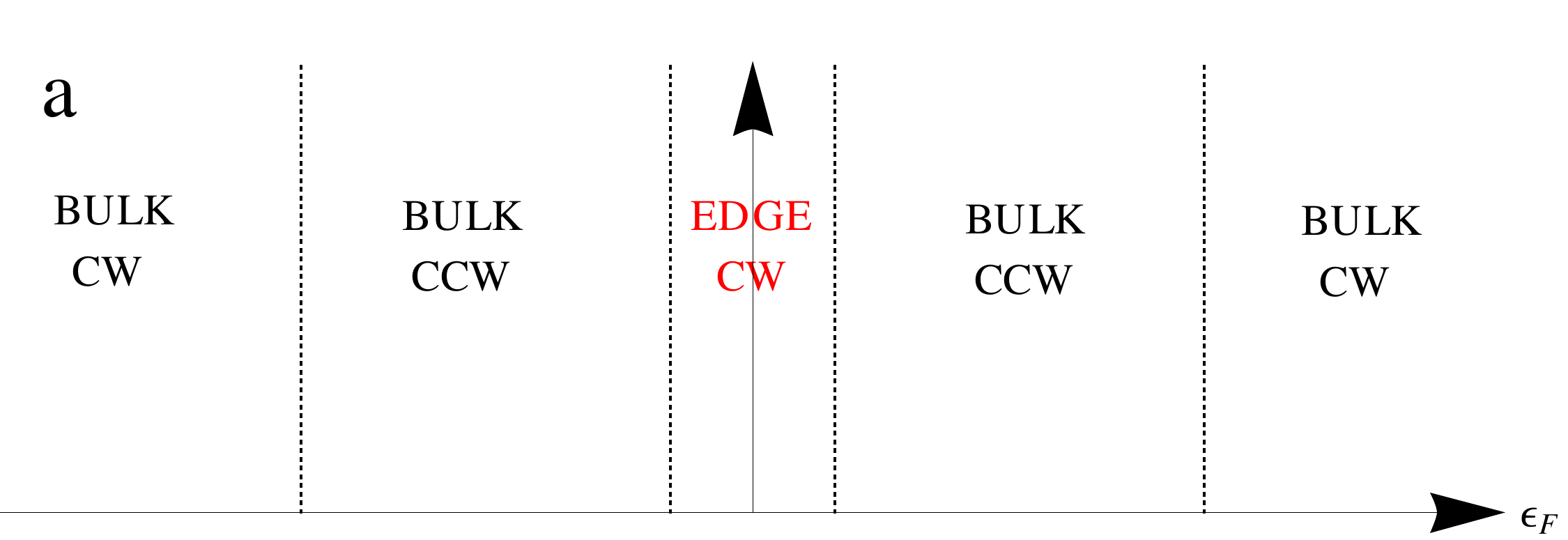}
\includegraphics[width=0.45\columnwidth]{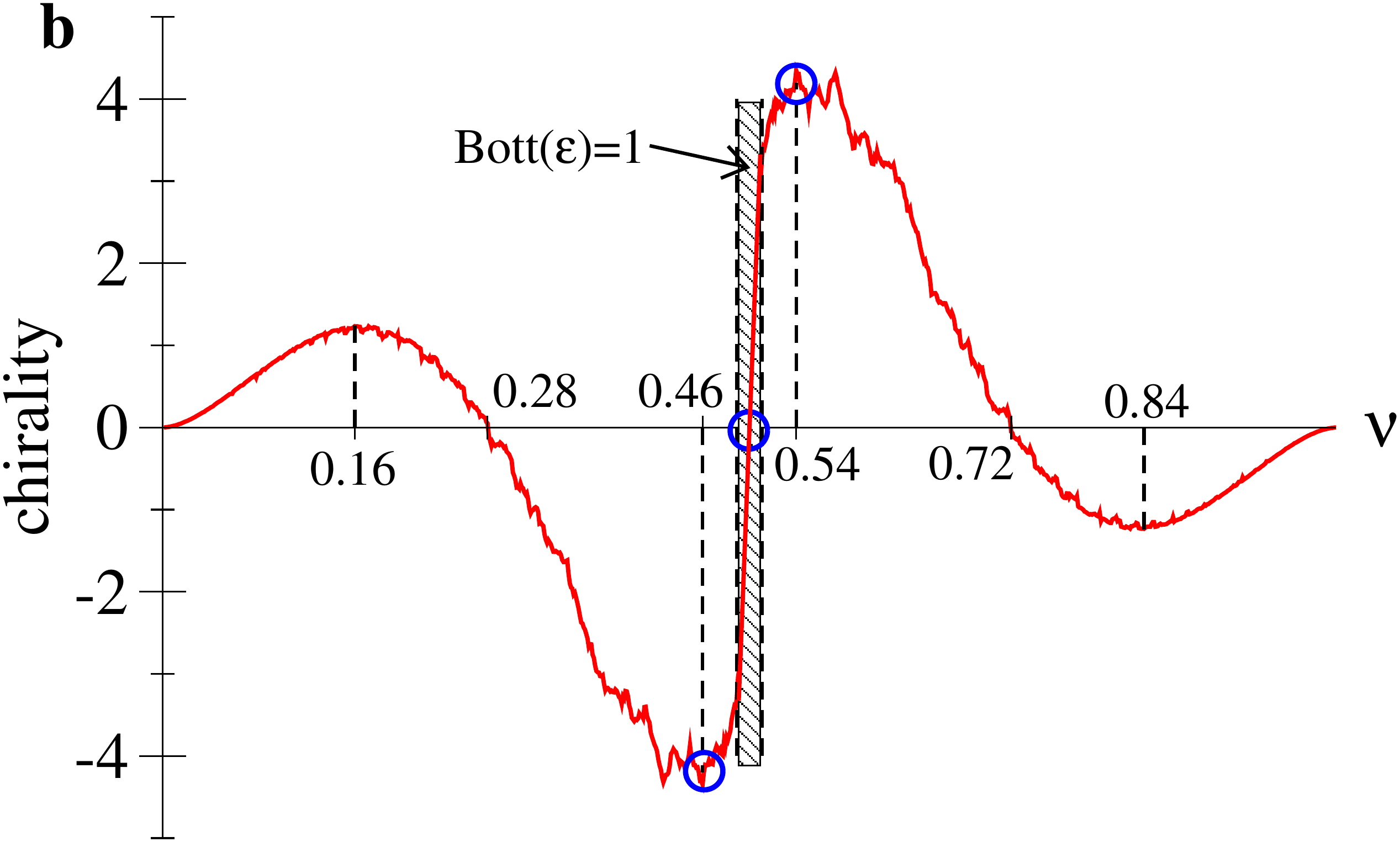}}
\captionof{figure}{ Time-averaged physical current in the Floquet 
eigenstates and Floquet Fermi Sea. {\bf{a}}. The eigenstates of $H_F$ that 
support the same chirality tend to be close in Floquet energy. CW stand for clockwise and CCW 
stand for counter-clockwise. {\bf{b}}. Chirality of the current in the Floquet Fermi sea changes 
with the filling fraction, and vanishes at $\nu=0.5$ and $\nu\approx 0.28, 0.72$. We have defined 
the chiral current as the sum of: i) the current moving to the right on the top edge, ii) the 
current moving downward on the right edge, iii) the current moving to the left on the bottom 
edge, and iv) the current moving upward on the left edge. Therefore a positive and negative values 
indicate clockwise and counter-clockwise edge current, respectively. The shaded box indicate 
the narrow interval of filling fraction ($0.49\le\nu\le0.51$) for which the equilibrium Bott 
index of the Floquet ground state is one.\label{fig:current-schematic}}
\end{minipage}

\begin{minipage}{\linewidth}% to keep image and caption on one page
\makebox[\linewidth]{%  
\includegraphics[width=0.30\columnwidth]{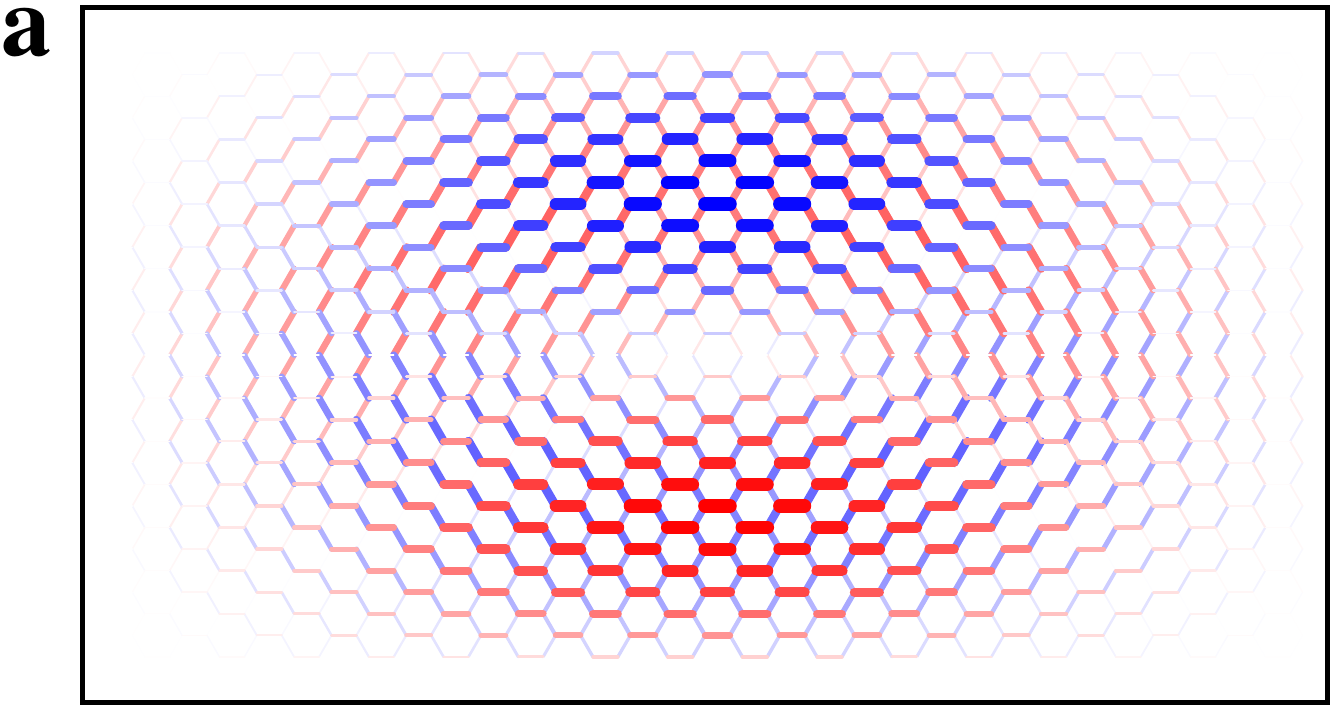}
\includegraphics[width=0.30\columnwidth]{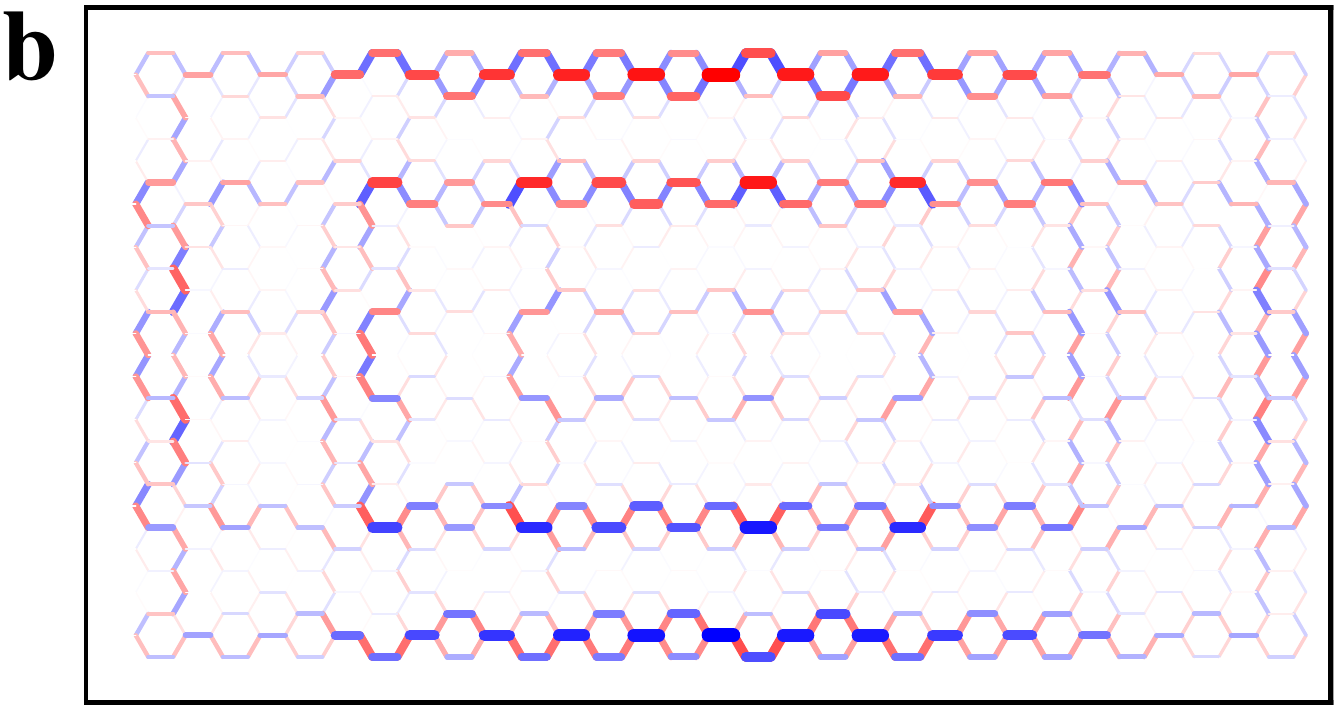}
\includegraphics[width=0.30\columnwidth]{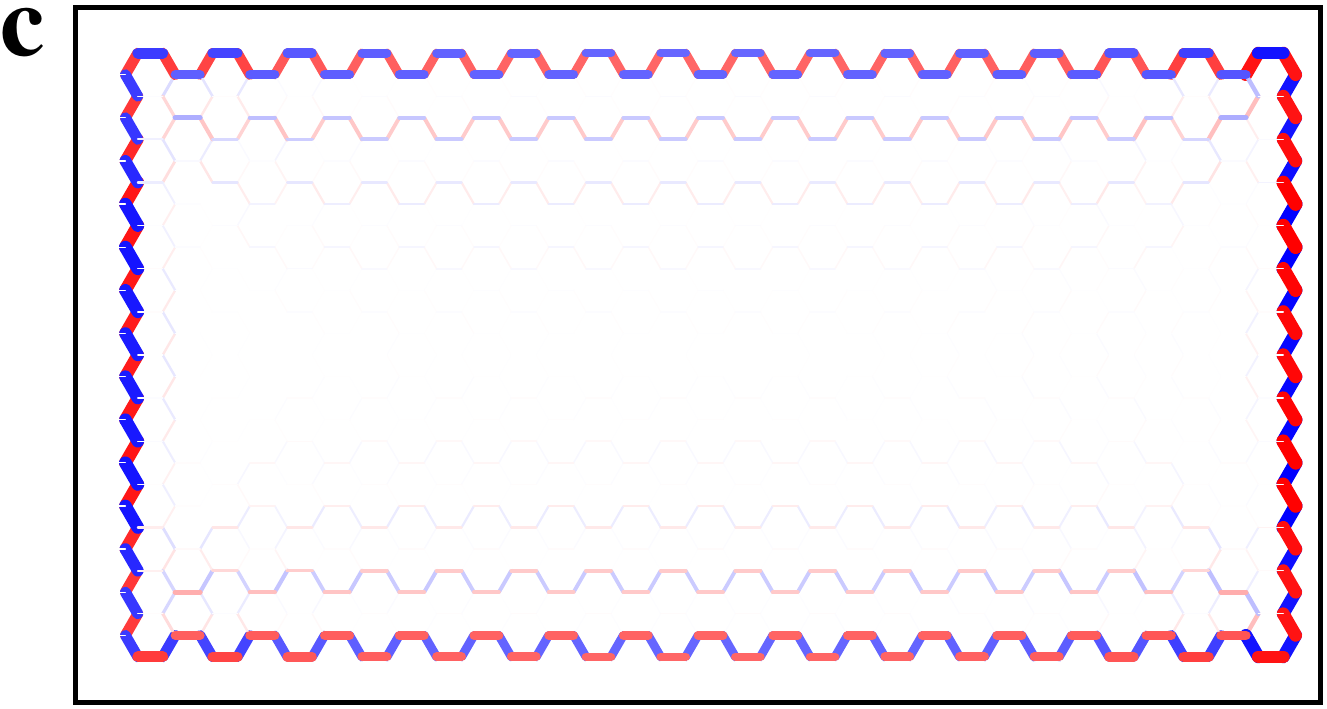}}
\captionof{figure}{ Time-averaged physical current in three 
Floquet eigenstates for a system in which the amplitude of the electric field is $E=7.0$. The eigenstates are: $\varepsilon=-2.28J,\, \text{state number}=1$ 
and  $\varepsilon=-0.79J,\,\text{state number}=321$, and  $\varepsilon=-0.01J,\, \text{state number}=464$ for panels {\bf{a}}, {\bf{b}} and {\bf{c}} respectively.  {\bf{a}}. The Floquet ground 
state supports a bulk current with clockwise chirality. {\bf{b}}. This excited 
eigenstate supports a counter-clockwise current with both an edge and a bulk component. 
{\bf{c}}. This topologically protected edge mode supports a clockwise edge current.\label{fig:eigen_eq_current}}
\end{minipage}

\begin{minipage}{\linewidth}% to keep image and caption on one page
\makebox[\linewidth]{%  
\includegraphics[width=0.30\columnwidth]{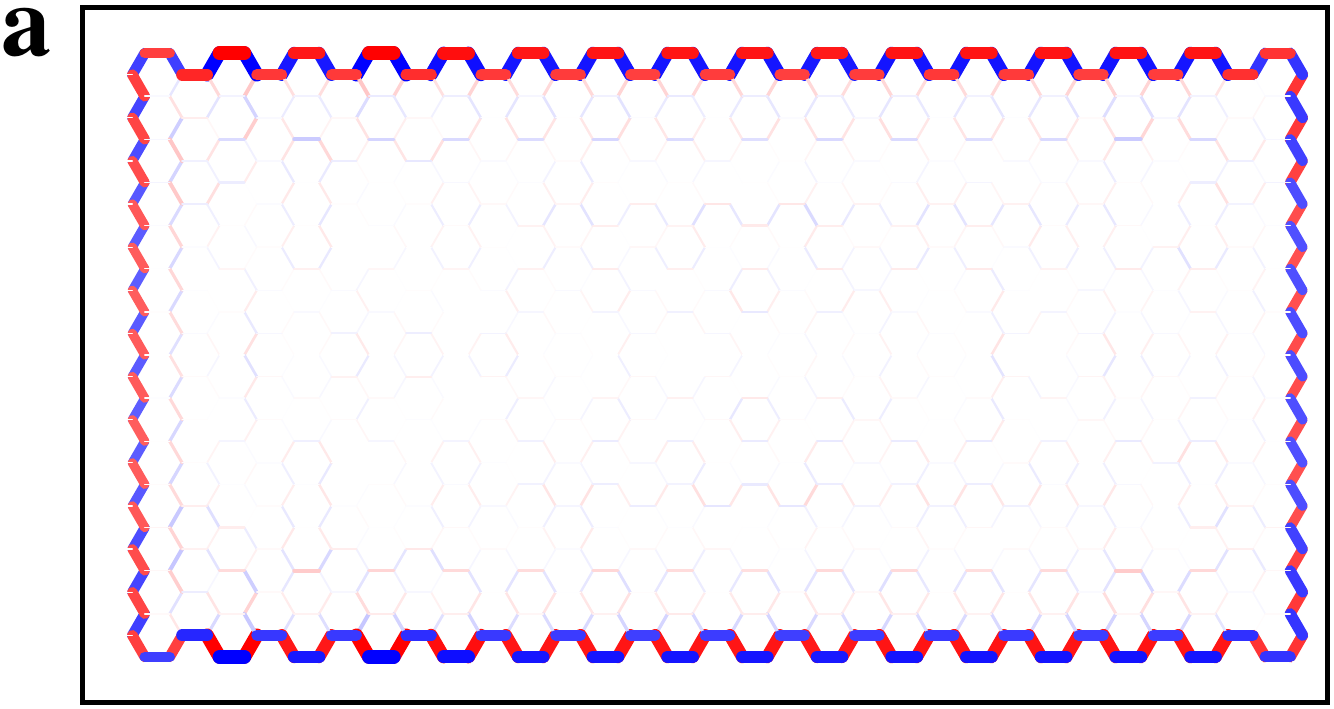}
\includegraphics[width=0.30\columnwidth]{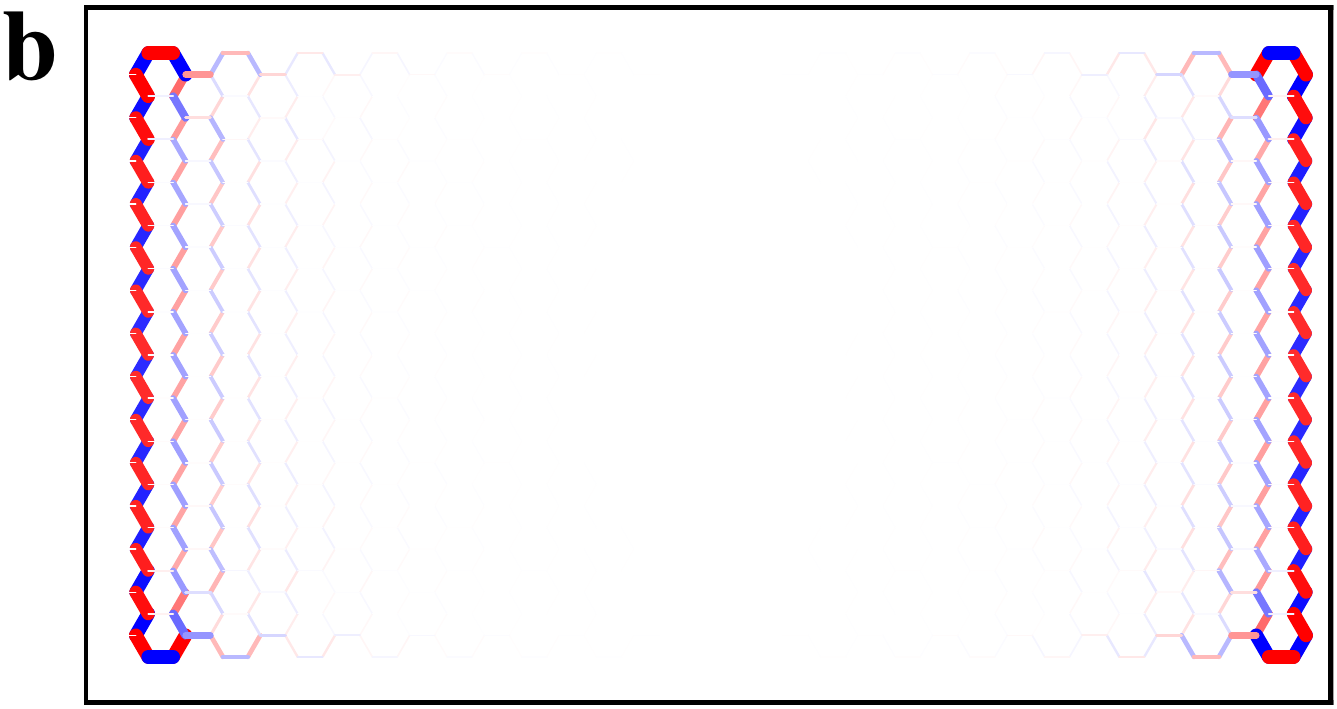}
\includegraphics[width=0.30\columnwidth]{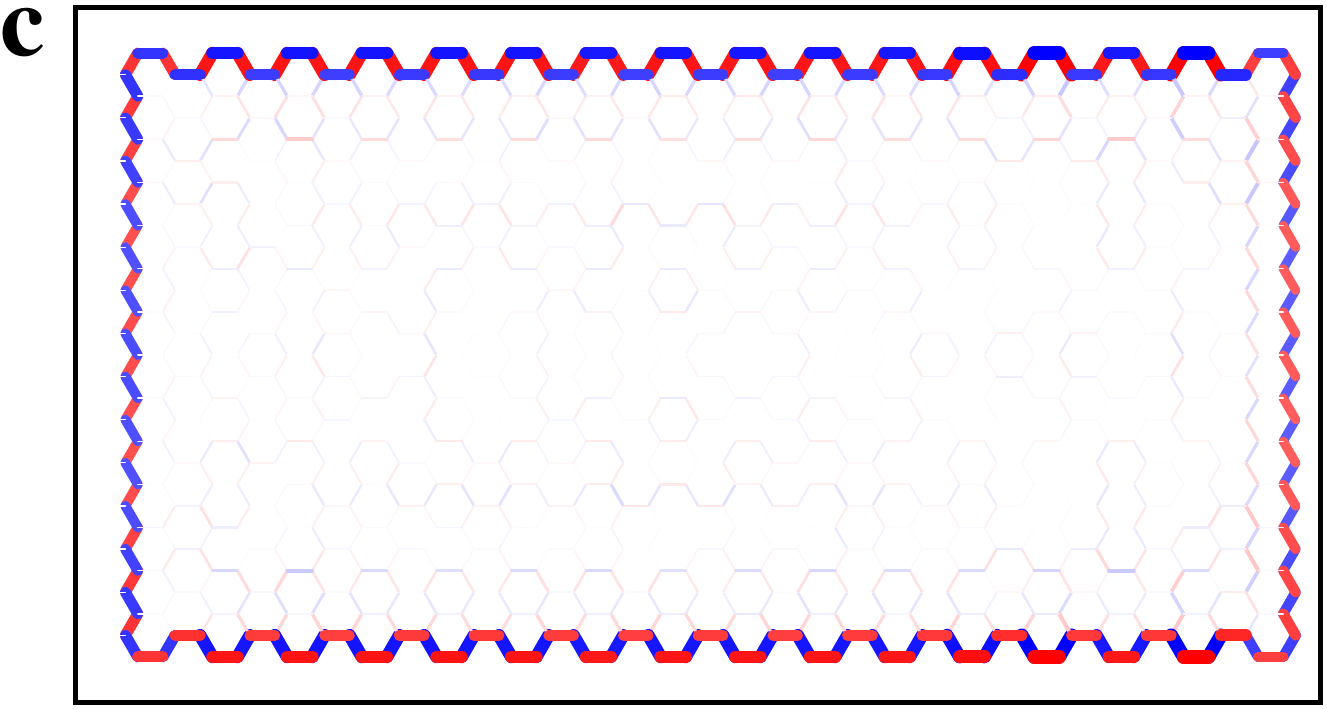}}
\captionof{figure}{ Time-averaged physical current in the exact Floquet 
Fermi sea for three values of the filling for a system 
in which the amplitude of the electric field is $E=7.0$. The fillings are: $\nu=0.46, 0.5$ and $0.54$ for panels {\bf{a}}, {\bf{b}} and {\bf{c}} respectively. 
{\bf{a}}. The current is localized on the edge and moves counter-clockwise. 
{\bf{c}}. The current is localized on the edge and moves clockwise. 
{\bf{b}}. At half-filling both chirality are presents. The current on the left and right 
edges moves downward and is compensated by a bulk current moving upward.\label{fig:GS_eq_current}} 
\end{minipage}

\begin{minipage}{\linewidth}% to keep image and caption on one page
\makebox[\linewidth]{%  
\includegraphics[width=0.30\columnwidth]{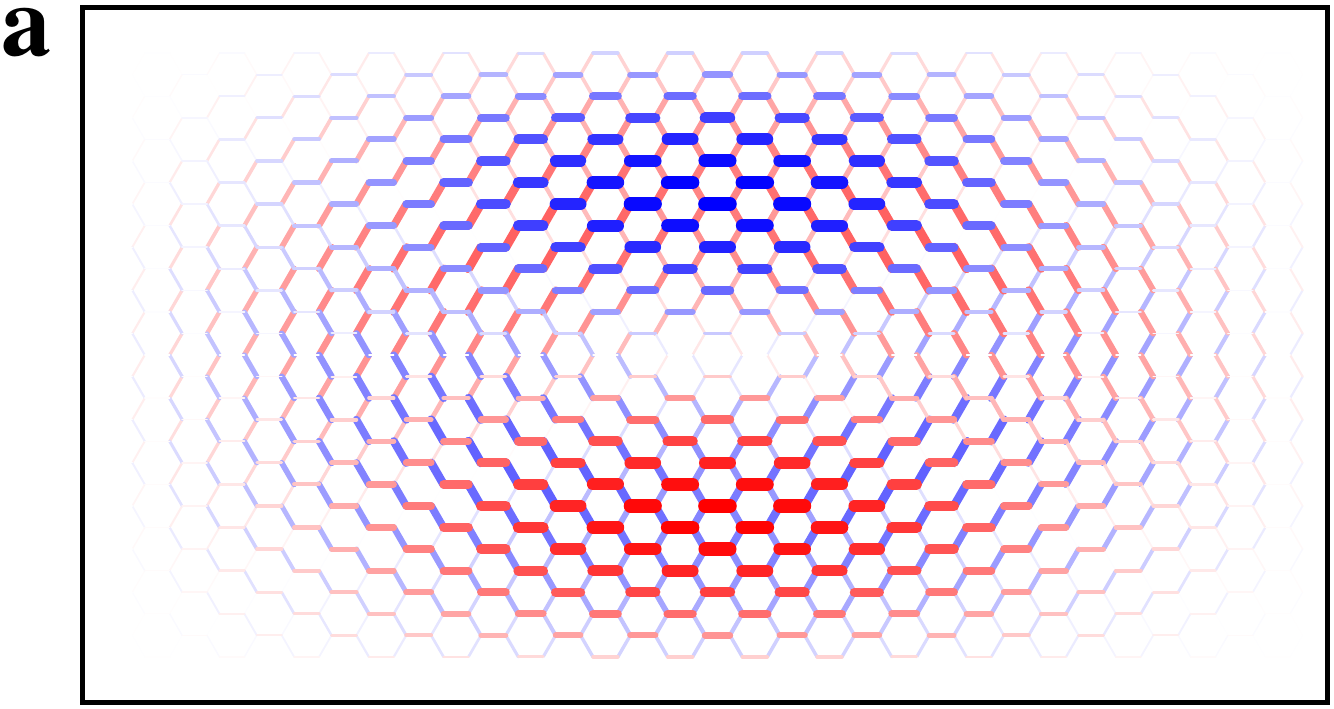}
\includegraphics[width=0.30\columnwidth]{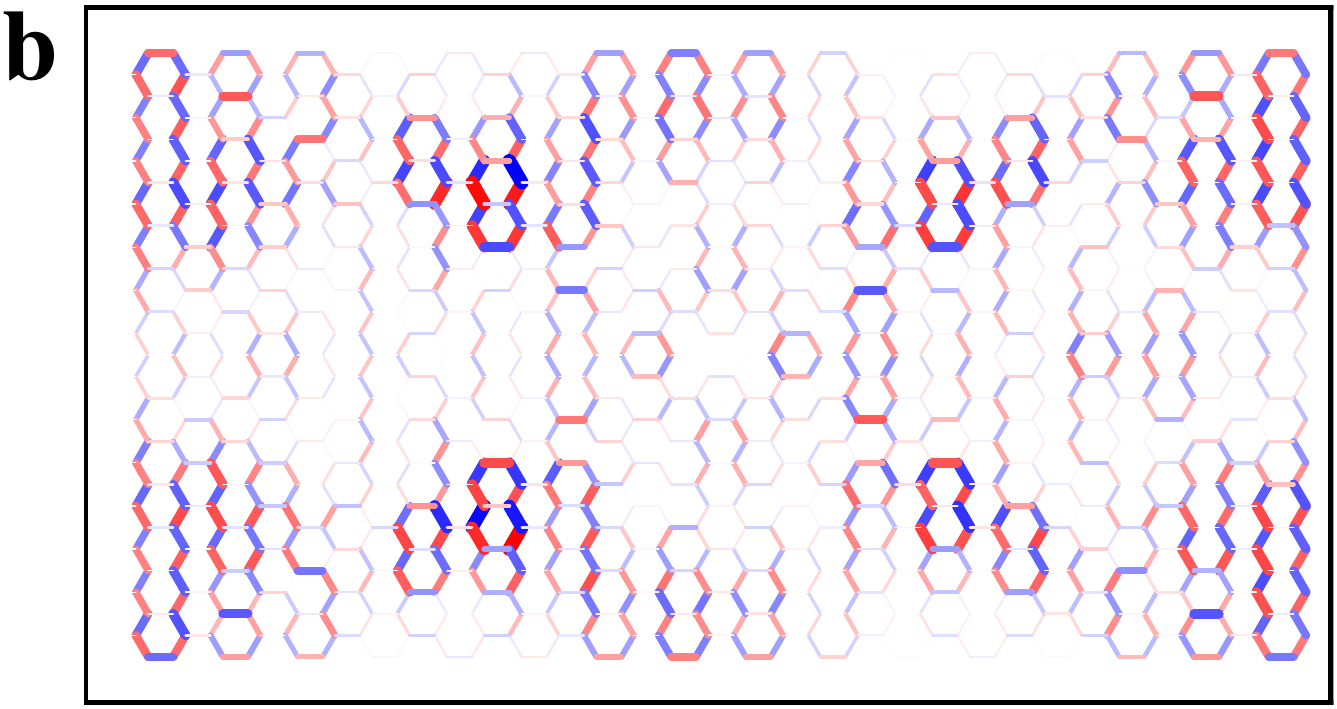}
\includegraphics[width=0.30\columnwidth]{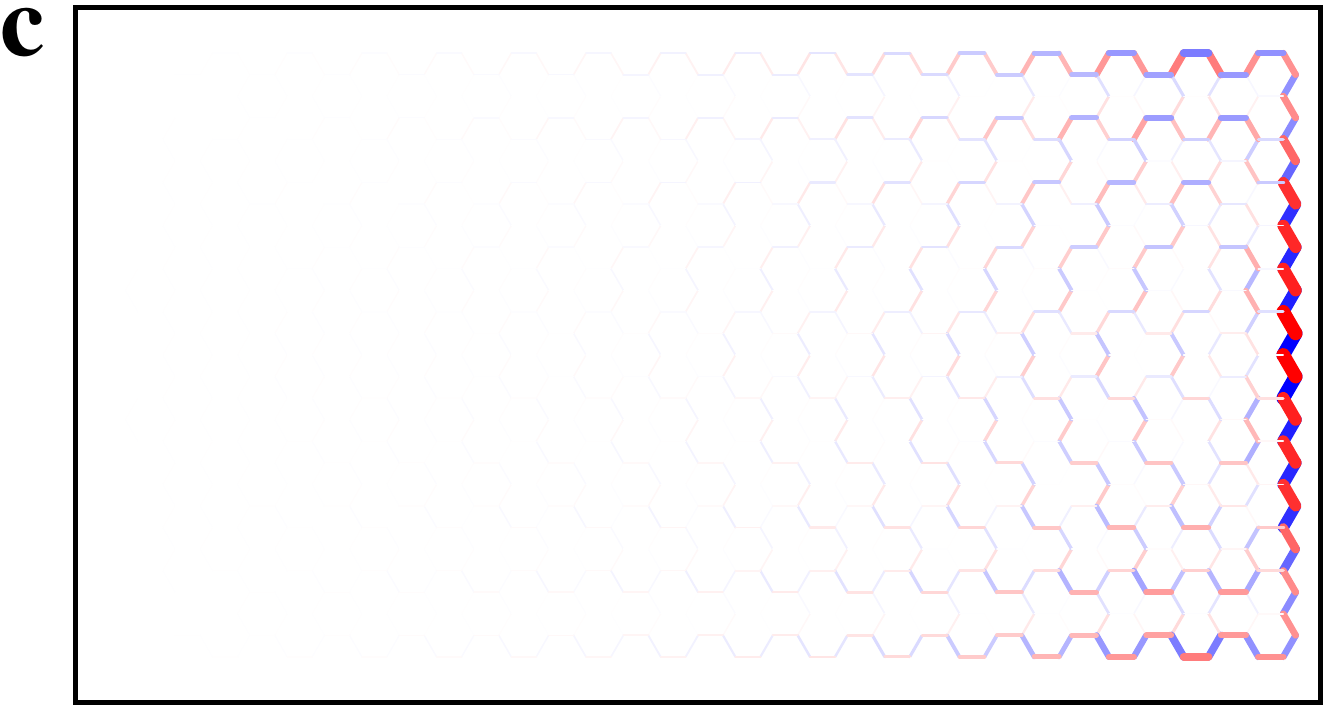}}
\captionof{figure}{Time-averaged physical current in three 
Floquet eigenstates for a system in which the amplitude of the electric field is $E=2.0$. 
The Floquet eigenstates are: {\bf{a}}. $\varepsilon=-2.93J\,\text{state number}=1$, 
{\bf{b}}. $\varepsilon=-1.04J,\,\text{state number}=321$ and 
{\bf{c}}. $\varepsilon=-0.06J,\,\text{state number}=464$.
The value $E=2.0$ is smaller than the critical one ($E_c\approx 5$) at which the Floquet Hamiltonian 
becomes topological. As expected, we do not observe an edge mode when entering the bulk gap 
(see panel {\bf{c}}) but bulk states away from the middle of the spectrum still can support 
chiral currents. For example, in panel {\bf{a}} one can see that the single-particle Floquet ground 
state still supports a clockwise current.\label{fig:NT_eigen_eq_current} }
\end{minipage}

\begin{minipage}{\linewidth}% to keep image and caption on one page
\makebox[\linewidth]{%  
\includegraphics[width=0.30\columnwidth]{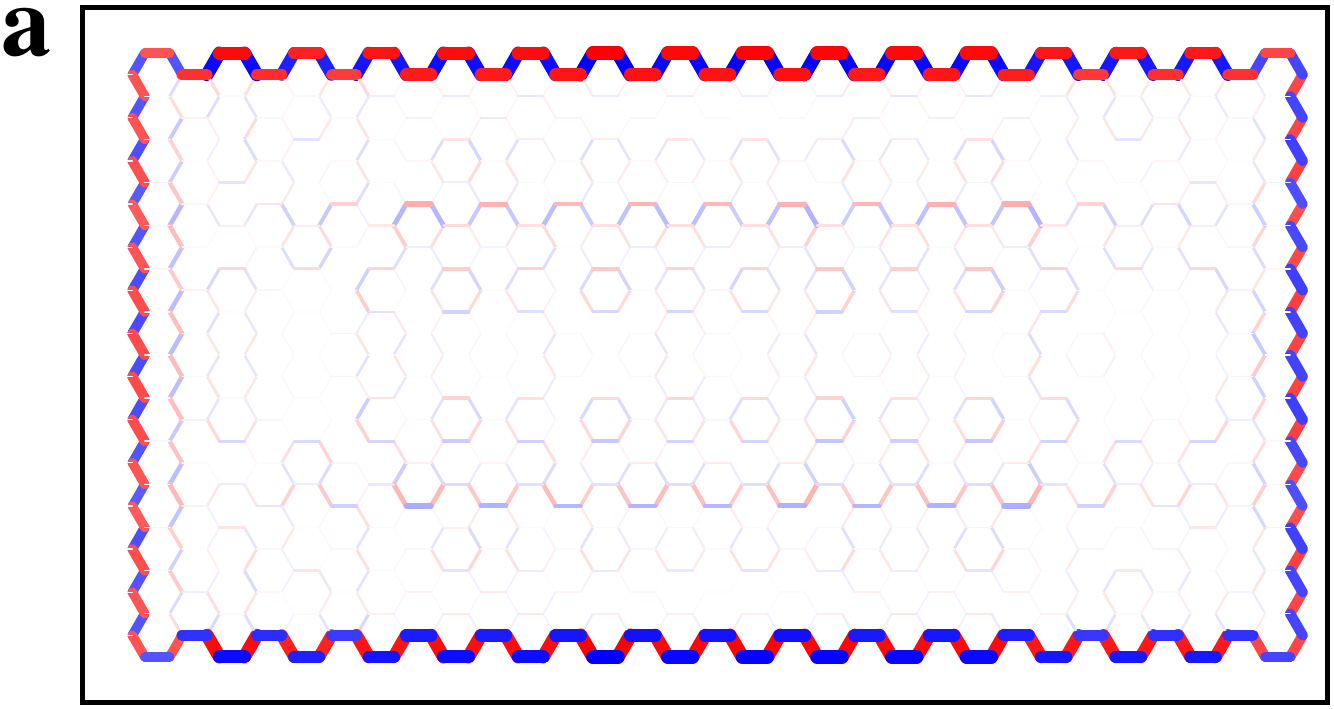}
\includegraphics[width=0.30\columnwidth]{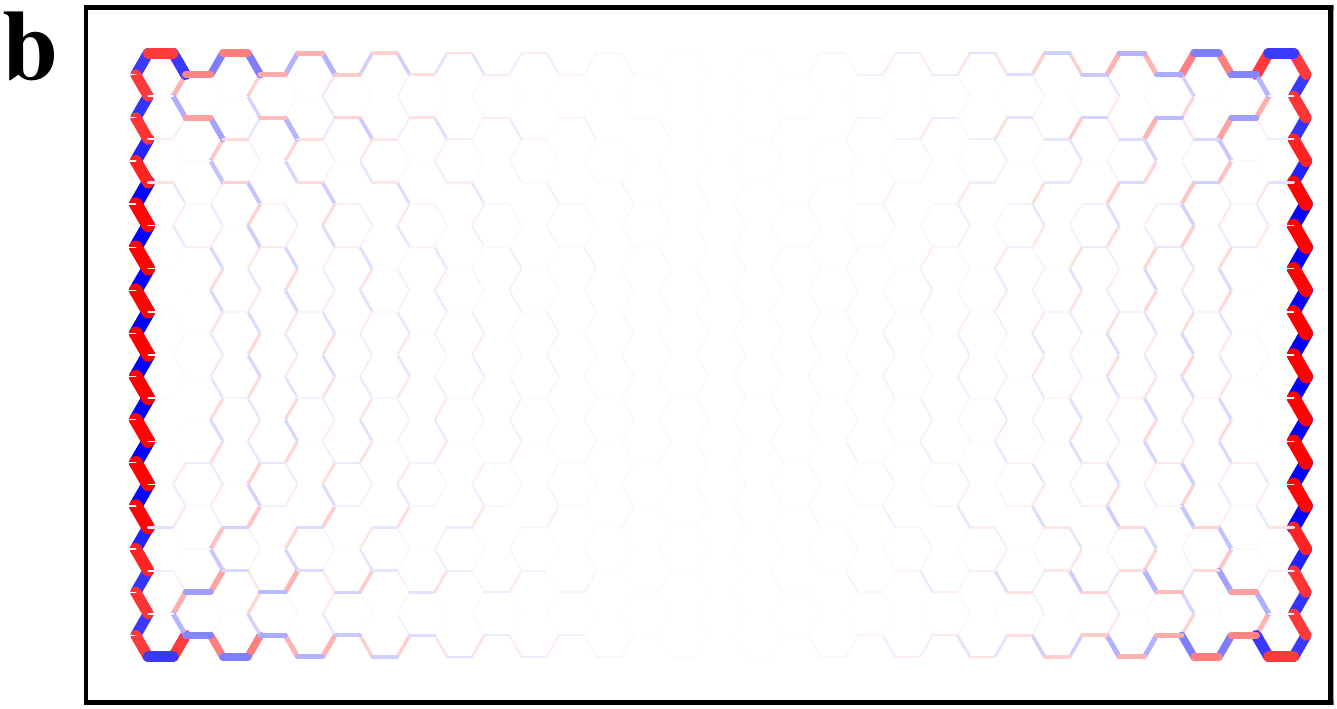}
\includegraphics[width=0.30\columnwidth]{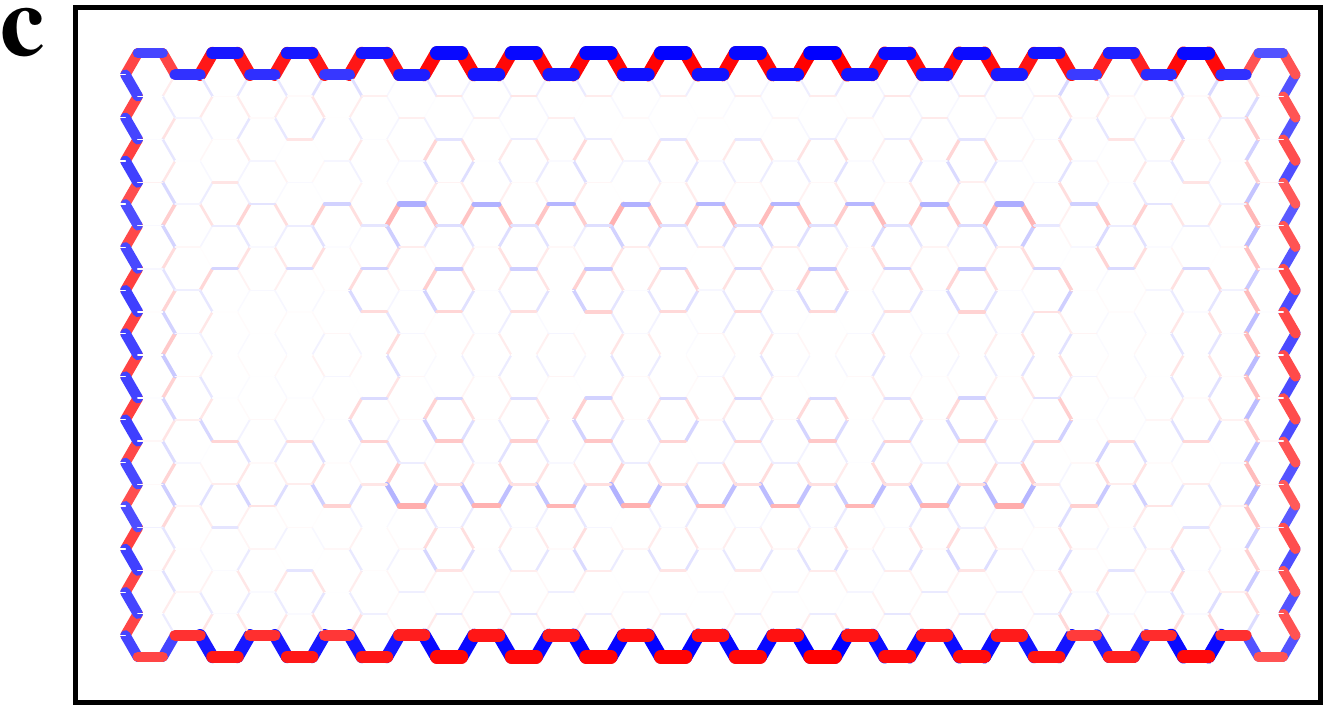}}
\captionof{figure}{Time-averaged physical current in the exact Floquet 
Fermi sea for three values of the filling for a system 
in which the amplitude of the electric field is $E=2.0$. 
The fillings are: $\nu=0.46, 0.5$ and $0.54$ for panels {\bf{a}}, {\bf{b}} and {\bf{c}} respectively. 
Despite the fact that $H_F$ is not topological, the current patter is similar to that seen in Supplementary Figure~\ref{fig:GS_eq_current}. Important quantitative differences between the these patters 
are discussed in the Supplementary Note~4.\label{fig:NT_GS_eq_current} }
\end{minipage}

\begin{minipage}{\linewidth}% to keep image and caption on one page
\makebox[\linewidth]{%  
\includegraphics[width=0.40\columnwidth]{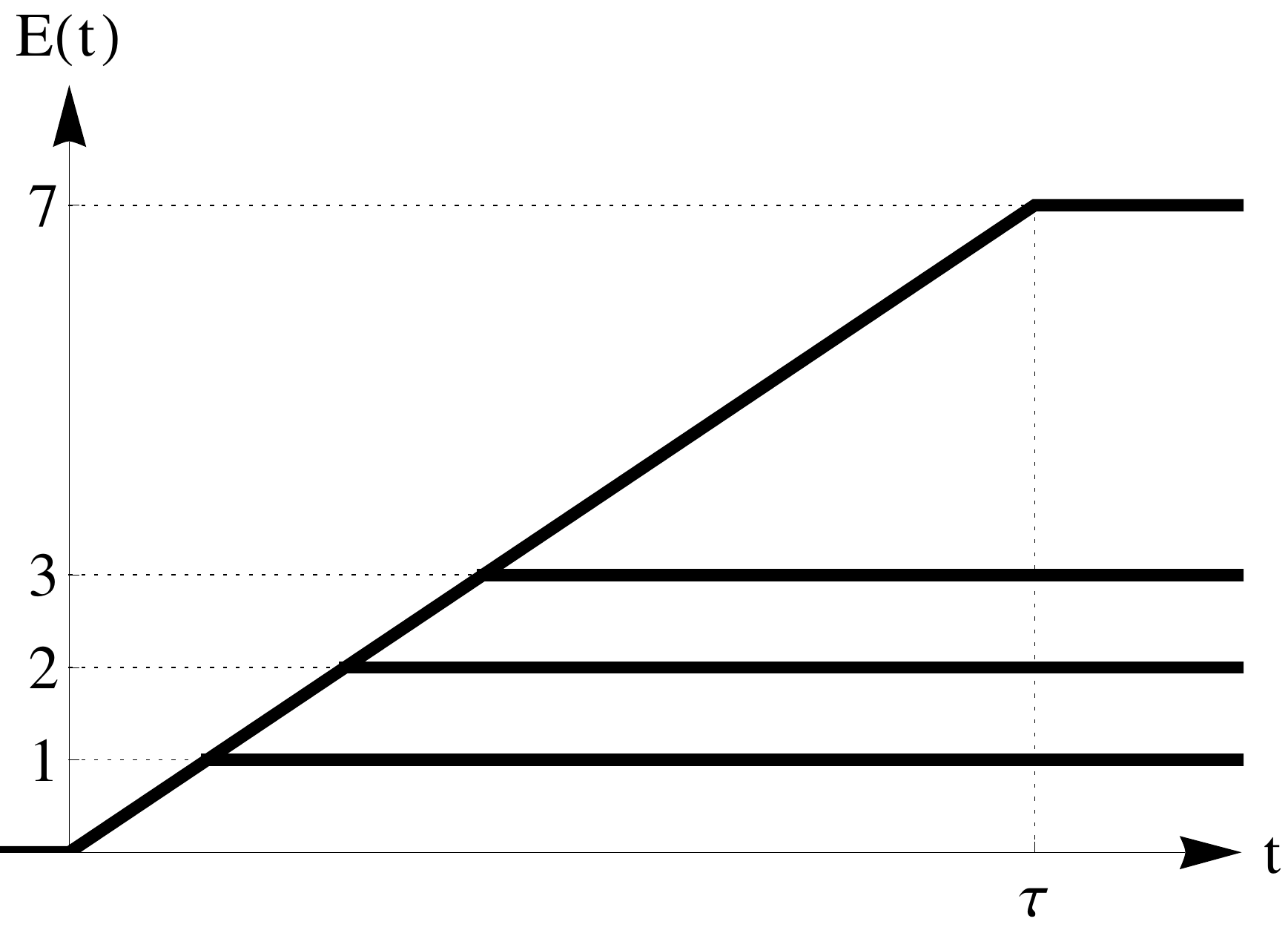}}
\captionof{figure}{\label{fig:scematic-many-ramps} 
Schematic picture of the four different ramp considered. Each ramp stops at a different value of the final electric field, $E_0=1,2,3,7$. Here $\tau=80 T$ as in the main text.} 
\end{minipage}

\begin{minipage}{\linewidth}% to keep image and caption on one page
\makebox[\linewidth]{%  
\includegraphics[width=0.90\columnwidth]{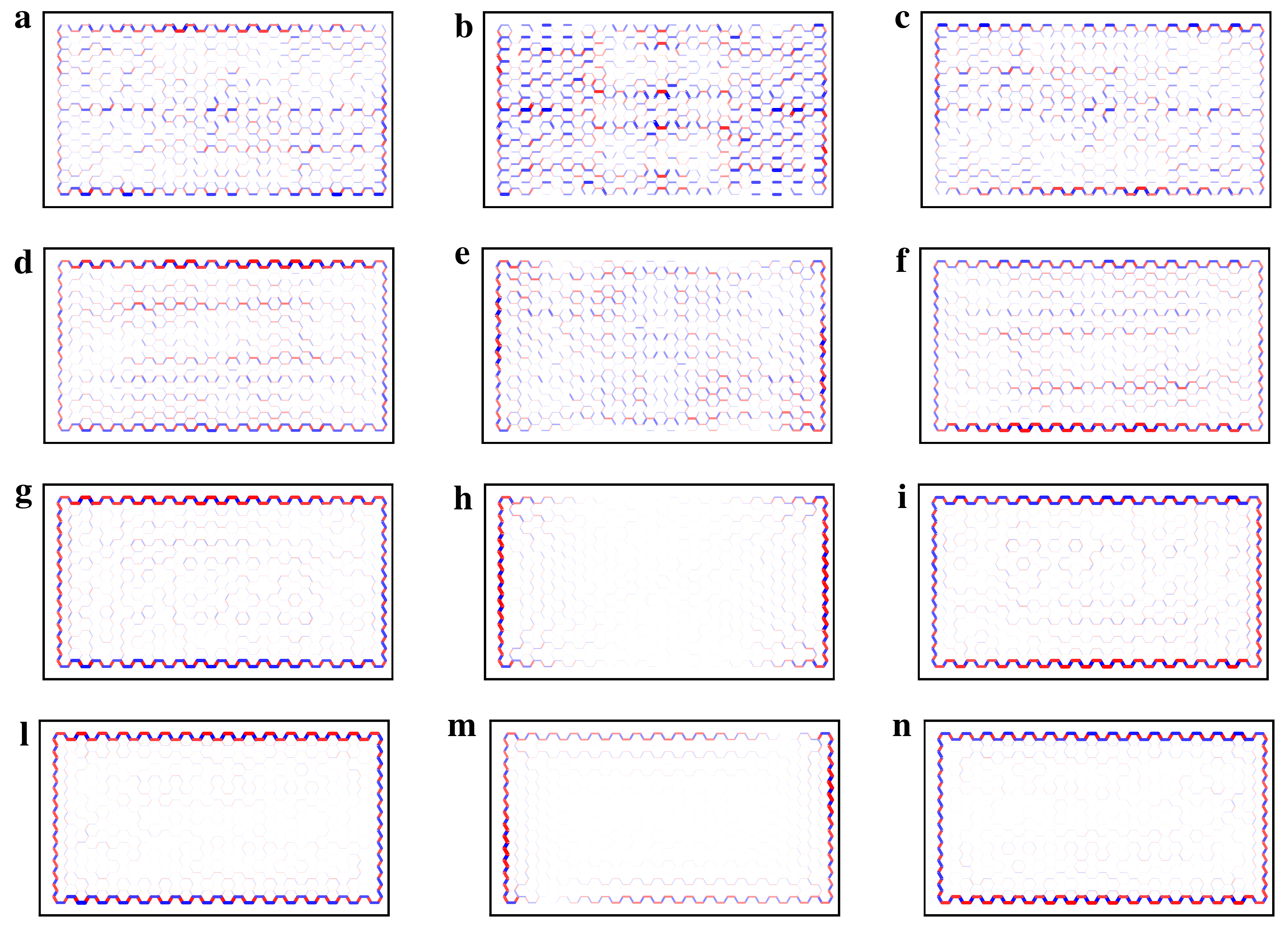}}
\captionof{figure}{\label{fig:NEQpanel} 
Time-averaged physical current at the end of different dynamical ramps. Each row corresponds to a different value of the final electric field $E_0=1,2,3,7$ (from top to bottom) and each column corresponds to a different value of the filling fraction $\nu=0.46, 0.5, 0.54$ (from left to right). The last row is identical to Fig.~\ref{fig:non_eq_current} in the main text and it is reported here for sake of comparison.} 
\end{minipage}

\begin{minipage}{\linewidth}% to keep image and caption on one page
\makebox[\linewidth]{%  
\includegraphics[width=0.40\columnwidth]{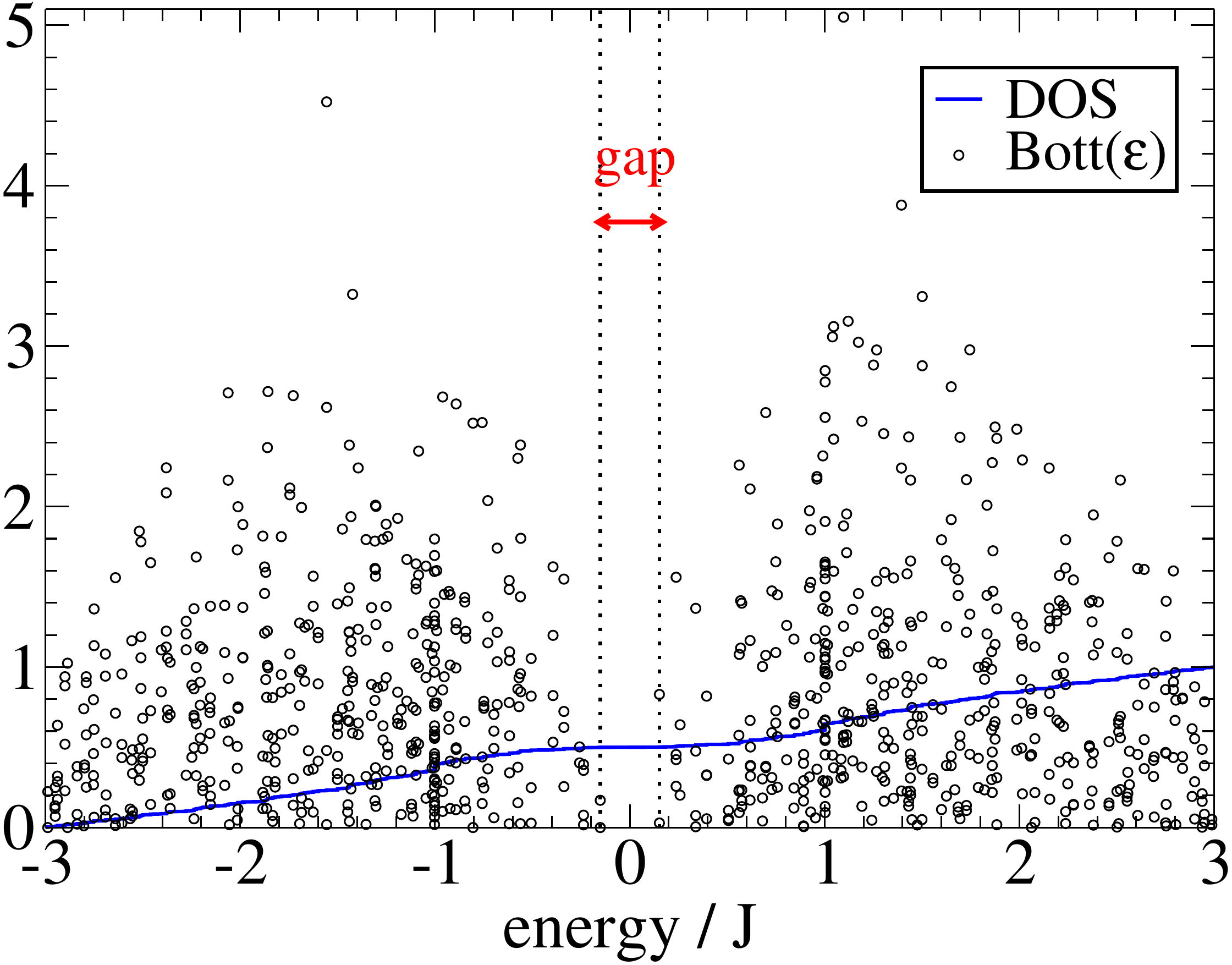}}
\captionof{figure}{\label{fig:bott_PBC} 
The density of states (DOS) and the equilibrium Bott index in a system with the static Hamiltonian $H_S$, with periodic boundary 
conditions.} 
\end{minipage}

%\newpage
%\section*{Supplementary Tables}
\pagestyle{empty}

\begin{table}[!h]
\begin{tabular}{|c|c|c|c|}
\hline 
 & $\nu=0.46$ & $\nu=0.5$ & $\nu=5.4$ \tabularnewline
\hline 
$E_0=1$ & $0.21$  & $0.09$ & $0.21$ \tabularnewline
\hline 
$E_0=2$ & $0.74$ &  $0.32$ & $0.74$ \tabularnewline
\hline 
$E_0=3$ & $1.17$ &  $0.52$ & $1.17$ \tabularnewline
\hline 
$E_0=7$ & $4.32$ & $1.00$ & $4.32$ \tabularnewline
\hline 
\end{tabular}
\caption{\label{table:max_val} 
Maximum values (in arbitrary units) of the current represented in the twelve panels in Supplementary Figure~\ref{fig:NEQpanel}.} 
\end{table}

%\section*{Supplementary Notes}

\subsection*{\label{sub:Appendix.specific-mapping} Supplementary Note 1: Explicit form of the 
pseudo-magnetic field $\mathbf{B}_{\mathbf{k}}(t)$}

Here, we report the explicit form of the pseudo-magnetic field $\mathbf{B}_{\mathbf{k}}(t)$ 
[see Eq.~\eqref{eq:mapping}] in the case of graphene subject to a circularly polarized 
uniform electric field. In the electromagnetic gauge, in which the electric field is 
represented via the vector potential, i.e., $\mathbf{E}(t)=-\partial_{t}\mathbf{A}(t)$, 
the Hamiltonian is block-diagonal in momentum space [see Eq.~\eqref{eq:mapping}]: 
\be
H_{\mathbf{k}}(t)=\left(c_{\mathbf{k},1}^{\dagger},
c_{\mathbf{k},2}^{\dagger}\right)\left(\begin{array}{cc}
\Delta/2 & -\overline{g\left(\mathbf{k},t\right)}\\
-g\left(\mathbf{k},t\right) & -\Delta/2
\end{array}\right)\left(\begin{array}{c}
c_{\mathbf{k},1}\\ c_{\mathbf{k},2}
\end{array}\right)=-\frac{1}{2}\,\mathbf{B}_{\mathbf{k}}(t)
\cdot \mathbf{\sigma}_{\mathbf{k}}, \label{def}
\ee
where the notation is as in Eq.~\eqref{eq:H} and the overline indicates complex conjugation. 
The function $g\left(\mathbf{k},t\right)$ describes the hopping between nearest neighbor 
lattice sites:
\be
g(\mathbf{k},t)\equiv\sum_{l=1}^{3}\,\, J_{l}(t)\,\, 
e^{i\mathbf{k}\cdot\mathbf{a}_{l}},\quad J_{l}(t)=
Je^{i\frac{e}{\hbar}\mathbf{A}(t)\cdot\mathbf{\delta}_{l}}.\label{def_g}
\ee
The lattice vectors $\mathbf{a}_{l}$ and nearest neighbor vectors 
$\mathbf{\delta}_{l}$ are:
\begin{align*}
\mathbf{a}_{1}&=\frac{a}{2}\left(3,\sqrt{3}\right), & 
\mathbf{a}_{2}&=\frac{a}{2}\left(3,-\sqrt{3}\right), & 
\mathbf{a}_{3}&=\left(0,0\right), \\
\mathbf{\delta}_{1}&=\frac{a}{2}\left(1,\sqrt{3}\right), & 
\mathbf{\delta}_{2}&=\frac{a}{2}\left(1,-\sqrt{3}\right), & 
\mathbf{\delta}_{3}&=a\left(-1,0\right),
\end{align*}
and the vector potential (when the electric field is fully on) is:
\be
\mathbf{A}(t)=\frac{E_{0}}{\Omega}\,
\left[\sin\left(\Omega\, t\right)\,,\,\cos\left(\Omega\, t\right)\right].
\ee
Putting all together, we obtain:
\be
\begin{split}
\frac{g(\mathbf{k},t)}{J}&=\exp\left[\frac{i}{2}\left(3k_{x}+
\frac{e}{\hbar}A_{x}(t)\right)+\frac{i\sqrt{3}}{2}\left(k_{y}+
\frac{e}{\hbar}A_{y}(t)\right)\right] \\
&+\exp\left[\frac{i}{2}\left(3k_{x}+\frac{e}{\hbar}A_{x}(t)\right)-
\frac{i\sqrt{3}}{2}\left(k_{y}+\frac{e}{\hbar}A_{y}(t)\right)\right]+
\exp\left[-i\frac{e}{\hbar}A_{x}(t)\right]. 
\label{g(k,t)}
\end{split}
\ee
It is immediate to verify that, for $\mathbf{A}=0$, the expression above reduces to:
\be
\frac{g(\mathbf{k})}{J}=1+\exp\left[i\left(\frac{3}{2}k_{x}+
\frac{\sqrt{3}}{2}k_{y}\right)\right]+\exp\left[i\left(\frac{3}{2}k_{x}-
\frac{\sqrt{3}}{2}k_{y}\right)\right],
\ee
which describes the band structure of undriven graphene and $J$ is the bare hopping. 
Note that both vectors $\mathbf{a}_{l}$ and $\mathbf{\delta}_{l}$ enter 
into the definition of $g(\mathbf{k},t)$ while only the vectors $\mathbf{a}_{l}$ 
enter into the definition of $g(\mathbf{k})$. For this reason, $g(\mathbf{k},t)$ 
is {\it not} obtained from $g(\mathbf{k})$ via Peierls substitution, i.e., 
$\mathbf{k}\rightarrow\mathbf{k}+\frac{e}{\hbar}\mathbf{A}(t)$. 

Given the form of the three Pauli matrices and Eq.~\eqref{def}, it is immediate to see that:
\be
B_{\mathbf{k}}^{x}(t)=2\,\text{Re}\left[g\left(\mathbf{k},t\right)\right],
\quad B_{\mathbf{k}}^{y}(t)=2\,\text{Im}\left[g\left(\mathbf{k},t\right)\right],
\quad B_{\mathbf{k}}^{z}(t)=-\Delta,
\ee
which, together with Eq.~\eqref{g(k,t)}, specifies the form of the time dependent 
pseudo-magnetic field $\mathbf{B}_{\mathbf{k}}(t)$ completely.

\subsection*{\label{SI:overlap} Supplementary Note 2: Overlap of the Fermi Sea of Static and Floquet Hamiltonians}

As described in the main text, the ground states $|\psi_0^{S}\rangle$ and $|\psi_0^{F}\rangle$
obtained by filling the valence bands of ${H}_{S}$ and ${H}_{F}$ have different topologies and are orthogonal to each other, see Eq.~\eqref{eq:overlap}. These two many-body states can be represented by the pseudo spin configurations $\mathbf{S}_{\mathbf{k}}^{S}$ and $\mathbf{S}_{\mathbf{k}}^{F}$, respectively.  
Despite having different topologies and being orthogonal to each other $|\psi_0^{S}\rangle$ has a large occupation on the valence band of ${H}_{F}$. 
This is because the pseudo spin configurations 
$\mathbf{S}_{\mathbf{k}}^{S}$ and $\mathbf{S}_{\mathbf{k}}^{F}$
differ significantly only close to the Dirac points, while
they are almost identical in the rest of the BZ. The latter can be seen in the 
momentum resolved overlap between $|\psi_0^{S}\rangle$ and $|\psi_0^{F}\rangle$
shown in Supplementary Figure~\ref{fig:BZ}{\bf{b}}. 
The normalized occupation (Occ) of the Floquet valence band is 
\be
\text{Occ}=\frac{1}{A_{BZ}}\iint_{BZ}\, dk_{x}dk_{y}\,
\left(\frac{1+\mathbf{S}_{\mathbf{k}}^{S}\cdot
\mathbf{S}_{\mathbf{k}}^{F}}{2}\right)\approx0.977,
\ee
where $A_{BZ}=\frac{8\pi^{2}}{3\sqrt{3}}$ is the area of the BZ.

\subsection*{\label{sub:Appendix.-edge-states} Supplementary Note 3: Edge States in the system with boundaries}

In Supplementary Figure~\ref{fig:edges-wf}, we show some single-particle Floquet eigenstates
at energies $\varepsilon\le0$ within and outside the bulk gap 
(see Fig. \ref{fig:boundary}\textbf{b} in the main text). As expected, the eigenstates with 
energies within the bulk gap are localized on the edges. It is interesting
to note that these states have much larger occupation on the zigzag
edges (left and right) than on the armchairs (top and bottom) edges.
The chiral nature of the edge states can be probed by preparing a single-particle
$\delta$-like wavefunction in the center of the left and right edge.
The subsequent time-evolution shows that part of the particle diffuses
towards the center of the sample while the rest remains localized
on the edges and moves clockwise. The fraction of charge that persists
on the edge is determined by the overlap between the initial $\delta$-like
wavefunction and the edge state and it is expected to scales as $1/L$
where $L$ is the length of the edge.

\subsection*{\label{sub:eq-currents} Supplementary Note 4: Currents in the exact Floquet Fermi sea for different values of the electric field and filling fraction}
\pagestyle{empty}

Here, we discuss in detail the physical currents in the exact topological Floquet Fermi seas when the 
amplitude of the electric field is large ($E_0=7$) so that the Floquet Hamiltonian is topological.
We stress that the physical current is obtained by evolving the Floquet Fermi Sea with the physical 
Hamiltonian $H(t)$ and it is different from the current obtained using $H_F$ (see Methods).
We find that the physical currents depend on the filling fraction (see Supplementary Figure~\ref{fig:current-schematic}). 
This is expected since at low fillings single-particle behavior needs to be recovered while 
at unit filling the current must vanish identically. The vanishing of the current at unit filling 
suggests the existence of bulk modes with a chirality opposite to that of the edge modes so that, 
when all modes are equally occupied, the net chirality (and current) is zero. Interestingly, 
we find that modes of equal chirality are close in Floquet energy. Obviously, near the center of 
the Floquet spectrum the clockwise topological protected edge modes are predominant. We also find 
clockwise (bulk) modes at very low and high Floquet energies. Finally, counter-clockwise bulk 
modes are observed just below and above the bulk gap (see 
Supplementary Figure~\ref{fig:current-schematic}{\bf{a}}). We expect the specifics of diagrams such as the 
one presented in Supplementary Figure~\ref{fig:current-schematic}{\bf{a}} to be model and parameter dependent. 
For our model and parameters studied, the edge current in the ground state of the Floquet 
Hamiltonian at half filling has no net chirality, a property that is shared by the time 
evolving state in Fig.~\ref{fig:non_eq_current} in the main text.

In Supplementary Figure~\ref{fig:eigen_eq_current}, we show the current in three individual Floquet 
eigenstates. Panel {\bf{a}} corresponds to the single-particle Floquet ground state, 
which supports a clockwise bulk current. Panel {\bf{b}} corresponds to a single-particle 
excited state, which supports a counter-clockwise currents which has both a edge and a bulk 
component. Panel {\bf{c}} corresponds to a single-particle eigenstate close to the center of 
the Floquet spectrum, i.e., which lies within the bulk gap, and therefore corresponds to a 
topologically protected edge mode. As expected, it supports an edge current with 
clockwise chirality.

In Supplementary Figure~\ref{fig:GS_eq_current}, we show the current in the Floquet Fermi sea for 
the three values of the filling, $\nu=0.46, 0.5,$ and 0.54. For $\nu=0.46$ and 0.54, 
the current is localized on the edges and reaches its maximum value. However, the edge 
current has opposite chiralities for $\nu<0.5$ and $\nu>0.5$ (see panels {\bf{a}} and {\bf{c}}). 
At exactly $\nu=0.5$ both chiralities are present making the net chirality of the current zero. 
In this case, the current on the left and right edges moves downward and is compensated by 
a bulk current moving upward. Therefore, close to the left edge we observe a counter-clockwise 
chirality while close to the right edge we observe a clockwise chirality. The symmetry 
between left and right edge is broken in our Hamiltonian by the sublattice potential.
The behavior shown here is reflected in the non-equilibrium current shown in Fig.~\ref{fig:non_eq_current} in the main text.

In Supplementary Figures~\ref{fig:NT_eigen_eq_current} and \ref{fig:NT_GS_eq_current}, we repeat the 
previous analysis when the amplitude of the electric field is $E_0=2$. This value is smaller 
than the critical value $E_c\approx 5$ at which the Floquet Hamiltonian becomes topological. 
A naive expectation would be that, in this case, we should not observe a current. However, this 
is not what happens. This is because the change of topology in the Floquet Hamiltonian 
correlates {\it only} with the presence/absence of an edge mode in the bulk gap. As one can see 
in Supplementary Figure~\ref{fig:NT_eigen_eq_current}, when the Floquet Hamiltonian is topologically trivial 
there is no edge mode as one enters the bulk gap (see panel {\bf{c}}), but the single particle 
Floquet ground state still supports a chiral bulk current (see panel {\bf{a}}). In 
Supplementary Figure~\ref{fig:NT_GS_eq_current}, we show the current in the Floquet Fermi Sea for the same 
three values of the filling as in Supplementary Figure~\ref{fig:GS_eq_current}. The current patterns in 
Supplementary Figures~\ref{fig:NT_GS_eq_current} and \ref{fig:GS_eq_current} look qualitatively the same. 
However, there are some important differences. First, the overall magnitude of the 
current in Supplementary Figure~\ref{fig:NT_GS_eq_current} is about $7$ times smaller than in 
Supplementary Figure~\ref{fig:GS_eq_current}. Moreover, at half filling, the current is localized close 
to the edge when the Floquet Hamiltonian is topological (see Supplementary Figure~\ref{fig:GS_eq_current}{\bf{b}}) while it extends more into the middle of the sample when the Floquet Hamiltonian is trivial 
(see Supplementary Figure~\ref{fig:NT_GS_eq_current}{\bf{b}}).

Summarizing, we find that the change in topology of the Floquet Hamiltonian {\it only} directly 
affects the edge modes in the bulk gap. However, the current in the many-particle Fermi sea has 
contributions from Floquet eigenstates with many quasi-energies that are not directly affected 
by the change in topology of $H_F$. Therefore, the signature of the topological nature of 
$H_F$ in the current is smeared out.

\subsection*{\label{sub:neq-currents} Supplementary Note 5: Non-equilibrium currents for different values of the final electric field and filling fraction}

Here, we compare the time-averaged physical currents at the end of different dynamical ramps.
In particular, we consider four ramps that stop at different values of the electric field $E_0=1,2,3,7$ (see Supplementary Figure~\ref{fig:scematic-many-ramps}). For each ramp, we consider three filling fractions: $\nu=0.46,\,0.5,\,0.54$. The corresponding currents are reported in the twelve panels in Supplementary Figure~\ref{fig:NEQpanel}.
As described in the main text, to obtain the time-averaged physical current we evolved the wave-function for $10^3$ periods after the electric field had reached its final value. Then, we compute the instantaneous current which is then averaged over a period. In Supplementary Figure~\ref{fig:NEQpanel} each row corresponds to a different value of the (final) electric field $E_0=1,2,3,7$ (from top to bottom) and each column corresponds to a different filling fraction $\nu=0.46, 0.5, 0.54$ (from left to right). The last row is identical to Supplementary Figure~\ref{fig:non_eq_current} in the main text and it is reported here for comparison. If the system was in the exact Floquet Fermi Sea the second row would be Supplementary Figure~\ref{fig:NT_GS_eq_current} and the fourth row would be Supplementary Figure~\ref{fig:GS_eq_current}.
We stress that only Supplementary Figure~\ref{fig:NEQpanel}{\bf{m}} corresponds to a topological wave-function. In fact, in the first three rows in Supplementary Figure~\ref{fig:NEQpanel} the electric field $E_0=1,2,3$ is smaller that the critical value $E_c\approx 5.0$  required to make the system topological, see Fig.~\ref{fig:EC(L,t)} in the main text. In the fourth column the electric field is larger than the critical value, i.e. $E_0=7>E_c$, so that the Floquet Hamiltonian is topological. However, only when the filling fraction is close to half, i.e. $0.49\le\nu\le5.1$, the wave-function can become topological (see Supplementary Figure~\ref{fig:current-schematic}{\bf{b}}). In order to produce these plots, we have normalized the current in each panel of Supplementary Figure~\ref{fig:NEQpanel} by its maximum value. These values are reported in Supplementary Table~\ref{table:max_val}. Due to this normalization procedure, the relative intensity of the current among different panels cannot be inferred from Supplementary Figure~\ref{fig:NEQpanel} and one has to rely on the values in Supplementary Table~\ref{table:max_val}.  
From Supplementary Figure~\ref{fig:NEQpanel} and Supplementary Table~\ref{table:max_val}, we see that by increasing the value of the final electric field for a fixed value of the filling fraction (i.e., moving from top to bottom along one of the column in Supplementary Figure~\ref{fig:NEQpanel} and Supplementary Table~\ref{table:max_val}) the time-averaged current becomes more localized along the edge of the sample and increases in intensity. We also note that the chirality of the edge current becomes more evident by increasing $E_0$. 
Finally we stress that, for each value of the electric field $E_0$, the current at $\nu=0.46$ and $\nu=0.54$ are related by spatial inversion with respect the center of the sample. For $\nu<0.5$ ($\nu>0.5$) the current has counter-clockwise (clockwise) chirality. Also, at $\nu=0.5$ the current is mapped into itself under spatial inversion with respect the center of the sample. This explain why, at half filling, both chiral component are observed.  

\thispagestyle{empty}
\subsection*{\label{sub:Bott_PBC} Supplementary Note 6: Bott index in clean systems with periodic boundary conditions}
\thispagestyle{empty}
In Supplementary Figure~\ref{fig:bott_PBC}, we show the the density of states (DOS) and the equilibrium Bott index in a clean system with periodic boundary conditions. 
The DOS has a zero slope about $\epsilon=0$ indicating the presence of a gap. 
The equilibrium Bott index is not quantized and it is also not a smooth function 
of the energy.

\end{widetext}
\end{document}